\documentclass[
floatfix,
noshowpacs,
preprintnumbers,
twocolumn,
aps,
superscriptaddress,
pra,
10pt,
nofootinbib,
longbibliography
]{revtex4-2}

\usepackage[T1]{fontenc}
\usepackage[utf8]{inputenc}
\usepackage{babel}

\usepackage{amsmath}
\usepackage{amssymb}
\usepackage{amsfonts}
\usepackage{mathtools}
\usepackage{physics}
\usepackage{braket}

\usepackage{bbm}
\usepackage{bbold}

\usepackage{graphicx}
\usepackage{dcolumn}
\usepackage{multirow}
\usepackage{booktabs}
\usepackage{tabularx}

\usepackage[usenames,dvipsnames]{xcolor}

\usepackage[colorlinks=true,linktocpage=true]{hyperref}
\hypersetup{
    allcolors={teal!70!black}
}

\newcommand{\HTot}{H_{\rm tot}}
\newcommand{\HS}{H_S}
\newcommand{\HP}{H_P}
\newcommand{\HSP}{H_{SP}}
\newcommand{\HB}{H_B}
\newcommand{\HSB}{H_{SB}}

\newcommand{\omz}{\omega_0}
\newcommand{\nup}{\nu}
\newcommand{\omp}{\omega_+}
\newcommand{\omm}{\omega_-}

\newcommand{\sigx}{\sigma_x}
\newcommand{\sigz}{\sigma_z}
\newcommand{\sigp}{\sigma_+}
\newcommand{\sigm}{\sigma_-}

\newcommand{\D}{\mathcal{D}}

\newcommand{\Jc}{\mathcal{J}_c}
\newcommand{\WP}{\mathcal{W}_P}

\renewcommand{\Tr}{\operatorname{Tr}}

\begin{document}


\title{Anti-Zeno boost in an autonomous quantized-piston thermal machine}

\author{M. Tahir Naseem}
\email{mnaseem16@ku.edu.tr}
\affiliation{Faculty of Basic Sciences, Ghulam Ishaq Khan Institute of Engineering Sciences and Technology, Topi 23640, District Swabi, Khyber Pakhtunkhwa, Pakistan}
\affiliation{National Center for Quantum Computing, University of Engineering and Technology Narowal, 10 km Muridke Road, Near Adda Siraj, Narowal, Pakistan}

\date{\today}

\begin{abstract}
Anti-Zeno enhancement of quantum heat machines has been established for working fluids driven by an external classical field,  but whether it can
enhance extractable-work accumulation in an autonomously evolving quantized piston remains an important question. Here we study an autonomous thermal machine in which a
two-level working fluid is coupled to a quantized harmonic piston and to two spectrally separated reservoirs, without external modulation. We derive a
finite-time master equation for the resulting carrier and sideband channels, whose rates interpolate between the Zeno, anti-Zeno, and Markovian regimes as
the coupling time is varied. For an initially coherent piston, the machine charges the piston and generates more ergotropy than the Markovian reference in
the anti-Zeno window. Reversing the same retained channels turns the device into a finite-resource refrigerator with enhanced cooling. The anti-Zeno
effect therefore increases the rate of operation while leaving the underlying carrier and sideband energy ratios unchanged. These channel ratios remain
consistent with the usual Carnot bounds over the operating regimes considered here. Our results establish finite-time reservoir sampling as a mechanism for
enhancing autonomous thermal-machine operation with a quantized piston.
\end{abstract}

\maketitle

\section{Introduction}
\label{sec:introduction}

Quantum thermal machines have become a central testbed for thermodynamics in the quantum regime~\cite{CANGEMI20241}. They extend the notions of work, heat,
and efficiency to systems built from only a few quantum degrees of freedom~\cite{VinjanampathyAnders2016}. Early studies established that a maser
can operate as a heat engine~\cite{Scovil1959,Geusic1967}, and this idea has since developed into a broader framework for quantum
thermodynamics~\cite{Kosloff2013,KosloffLevy2014,VinjanampathyAnders2016, GelbwaserAdv2015,MyersAbahDeffner2022,Naseem2024HeatCurrents,
CANGEMI20241,Campbell_2026}. Quantum thermal machines are commonly classified as reciprocating or continuous. In reciprocating machines, a working medium
undergoes distinct strokes, whereas in continuous machines the relevant energy-exchange processes act
simultaneously~\cite{Kosloff2013,KosloffLevy2014,GelbwaserAdv2015, MyersAbahDeffner2022,Naseem2024HeatCurrents,CANGEMI20241,Campbell_2026,
Naseem:19,Naseem_2020}. Continuous operation can also be autonomous, with the machine evolving without externally timed modulation after initialization.
The present work considers such a continuous autonomous thermal machine with a quantized piston.

Many quantum thermal machines are described semiclassically, with a quantum working fluid exchanging energy with hot and cold reservoirs while a
prescribed time-dependent field or classical piston supplies or extracts work~\cite{KosloffLevy2014,GelbwaserKurizkiPRE2014}. Replacing this prescribed
drive with a dynamical quantum degree of freedom yields an autonomous machine without externally imposed time
dependence~\cite{YoussefMahler2010,GelbwaserKurizkiPRE2014}. In the minimal quantized-piston framework~\cite{GelbwaserPRE2013,GelbwaserKurizkiPRE2014},
a two-level working fluid is coupled to a harmonic piston and two reservoirs, with the piston acting as the work
repository~\cite{GelbwaserEPL2013,MariFarace2015,GelbwaserSciRep2015}. Related autonomous machines have been proposed using quantum
rotors~\cite{Roulet2018} and superconducting-circuit architectures~\cite{Rasola2025}. Because the piston is a dynamical quantum
subsystem, its mean energy need not be fully extractable. The ergotropy quantifies its useful work content as the maximum work extractable by a
cyclic unitary~\cite{Allahverdyan2004,PuszWoronowicz1978,Lenard1978}. The piston can therefore serve as a bosonic quantum battery charged by the
engine, with ergotropy quantifying the stored energy available for work extraction~\cite{AlickiFannes2013,Binder2015,Campaioli2024,Farina2019}.

Experimental advances increasingly provide access to quantum working media, reservoirs, and dynamical loads. Realizations include a single-ion heat
engine~\cite{Abah2012,Rossnagel2016}, continuous operation in a semiconductor quantum dot, and cyclic engines based on nitrogen-vacancy centers, nuclear
spins, and atomic collisions~\cite{Josefsson2018,Klatzow2019,Peterson2019, Bouton2021}. Particularly relevant here are a spin engine coupled to a
harmonic-oscillator flywheel~\cite{vonLindenfels2019} and an energy-conversion device whose quantum vibrational load was characterized through
ergotropy~\cite{VanHorne2020}. These experiments connect thermal-machine operation to the accumulation of extractable work in a quantum load.
Further developments include exceptional-point control~\cite{Zhang2022EP}, absorption refrigeration in trapped-ion and
autonomous superconducting-circuit settings~\cite{Maslennikov2019,Aamir2025}, a transmon-based Otto engine~\cite{Uusnakki2025}, and a recent work
reporting an autonomous superconducting heat engine~\cite{Uusnakki2026Autonomous}.

Beyond the amount of ergotropy stored in the piston, the rate at which it accumulates determines the useful charging performance. For a fixed initial
preparation, a greater ergotropy gain over the same operating interval corresponds to a higher average ergotropy charging power. Finite-time
operation is therefore central to this task, although it does not by itself invalidate a Markovian description.
Reservoir-memory corrections become important when the system--reservoir interaction time \(\tau_c\) is comparable to or shorter than the reservoir
correlation time \(\tau_B\). In this regime, the transition rates become sensitive to both elapsed time and local spectral structure. Such memory
effects can modify heat currents and engine performance~\cite{Thomas2018,Pezzutto2018,Wiedmann2020} and enhance the
maximum power of quantum thermal machines~\cite{AbiusoGiovannetti2019}. Complementarily, bath spectral filtering exploits reservoir spectral
structure to tailor energy exchange and quantum-state generation in open quantum systems~\cite{KofmanKurizkiSherman1994,NaseemPRR2020,
NaseemCommunPhys2021,NaseemPRA2022,NaseemQST2022}. Together, these developments motivate investigating whether finite-time
reservoir sampling can improve autonomous charging of a quantum piston.

Zeno and anti-Zeno dynamics provide a natural framework for addressing this question. Originally associated with repeatedly observed unstable systems,
the quantum Zeno effect suppresses decay under sufficiently frequent observations, whereas finite intervals can instead accelerate it through
the anti-Zeno effect~\cite{MisraSudarshan1977,Itano1990,KofmanKurizki2000,Fischer2001}.
More generally, finite-time evolution broadens the system response into a spectral kernel whose overlap with the reservoir determines whether a
transition rate is suppressed or enhanced relative to its long-time value~\cite{KofmanKurizki2000,KoshinoShimizu2005,Maniscalco2006}.
Here, anti-Zeno enhancement denotes a finite-time transition rate above its long-time golden-rule value, with no measurement protocol implied.
This mechanism has increased the power and cooling current of fast-driven heat machines~\cite{MukherjeeCommunPhys2020} and enhanced a driven minimal
machine in a structured bandgap environment~\cite{XuBandgap2022}. These studies employ externally prescribed modulation of the working-medium
transition frequency. Bringing this mechanism into an autonomous quantized-piston setting raises a further thermodynamic question.
Enhanced energy transfer can contribute both to extractable work and to passive energy as the piston state evolves. Enhanced transition rates
alone therefore do not establish improved useful output. The relevant question is whether finite-time reservoir response increases the
accumulation of piston ergotropy and how this benefit depends on the initial preparation.

In this work, we address this question for an autonomous machine comprising a two-level working fluid, a harmonic piston, and two spectrally separated
thermal reservoirs. We derive a finite-time master equation in the polaron frame that resolves the dynamics into carrier and sideband channels.
The corresponding rates capture the crossover from Zeno suppression through anti-Zeno enhancement to the long-time Markovian limit as the coupling time
is varied. For an initially coherent piston, finite-time reservoir sampling increases the ergotropy generated relative to the corresponding Markovian evolution.
The comparison with other piston preparations shows that this useful-work benefit is state dependent. Under the
refrigerator bias, the same retained channels yield enhanced cooling during consumption of the finite piston resource. The coherent-piston engine and
refrigerator benchmarks are confirmed by numerical solution of the full
joint working-fluid and piston dynamics. Finite-time sampling changes the transition rates without changing the
underlying channel energies. The corresponding channel-energy ratios remain Carnot
compatible over the regimes considered. Here the enhancement arises from the finite-time response of thermal reservoirs. This differs
from schemes using squeezed or other nonthermal reservoirs, whose additional resources can modify the applicable efficiency
bound~\cite{Rossnagel2014,Klaers2017,Niedenzu2016,Niedenzu2018}.

The paper is organized as follows. Section~\ref{sec:model_master_equation} formulates the model and derives the
finite-time master equation. Section~\ref{sec:quantum_heat_engine} treats the engine and the anti-Zeno enhancement of piston ergotropy.
Section~\ref{sec:quantum_refrigerator} treats the refrigerator and its finite-time enhancement. Section~\ref{sec:conclusion} concludes.
The appendices present the microscopic derivation of the master equation~(Appendix~\ref{app:finite_time_master_equation}), the state-resolved
ergotropy analysis~(Appendix~\ref{app:piston_state_ergotropy}), and the channel-energy and ergotropy-based performance measures~(Appendix~\ref{app:efficiency_cop}).

\section{Model and Master Equation}
\label{sec:model_master_equation}

We first formulate the microscopic model and derive the dynamical equation used throughout the paper. The working fluid is a two-level system (TLS) coupled to
a harmonic oscillator that acts as a quantized piston and to two independent thermal reservoirs, labelled hot and cold. At this stage, the reservoir response
functions are left arbitrary. In Secs.~\ref{sec:quantum_heat_engine} and \ref{sec:quantum_refrigerator}, we specify the reservoir spectra such that the
hot carrier and cold lower sideband constitute the retained channels for both operating modes.

\subsection{Microscopic setup}
\label{subsec:microscopic_setup}

\begin{figure*}[t]
    \centering
    \includegraphics[width=0.98\textwidth]{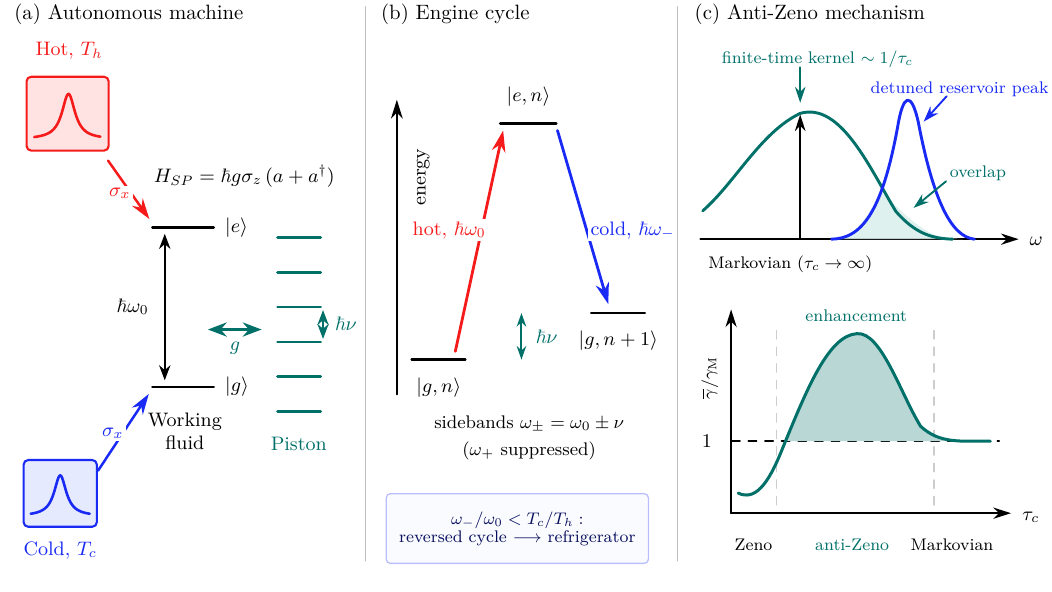}
\caption{
    Schematic of the autonomous quantum thermal machine.
    (a) A two-level working fluid, with states \(\ket{g}\) and \(\ket{e}\)
    separated by \(\hbar\omega_0\), is coupled longitudinally to a quantized
    harmonic piston of frequency \(\nu\) through
    \(H_{SP}=\hbar g\sigma_z(a+a^\dagger)\), and
    transversely through \(\sigma_x\) to hot and cold structured reservoirs at
    temperatures \(T_h\) and \(T_c\).
    (b) Joint working-fluid--piston energy levels in the engine regime. The hot
    reservoir drives the carrier transition
    \(\ket{g,n}\rightarrow\ket{e,n}\) at energy \(\hbar\omega_0\), while the
    cold reservoir drives the lower-sideband transition
    \(\ket{e,n}\rightarrow\ket{g,n+1}\) at energy \(\hbar\omega_-\). Each
    completed cycle adds one piston quantum \(\hbar\nu\), with
    \(\hbar\omega_0=\hbar\omega_-+\hbar\nu\). The sidebands are
    \(\omega_\pm=\omega_0\pm\nu\), and the upper sideband \(\omega_+\) is
    spectrally suppressed. Reversing the cycle produces refrigeration when
    \(\omega_-/\omega_0\lesssim T_c/T_h\)
    (Sec.~\ref{sec:quantum_refrigerator}).
    (c) Anti-Zeno mechanism. The finite-time sinc kernel entering
    Eq.~\eqref{eq:finite_time_rate_main}, with characteristic width
    \(\sim1/\tau_c\), can overlap a detuned reservoir peak more strongly than the
    narrow Markovian sampling obtained as \(\tau_c\rightarrow\infty\), thereby
    enhancing the transition rate. The lower diagram shows schematically the
    coupling-averaged finite-time rate normalized by its Markovian value, as in
    Eq.~\eqref{eq:AZD_channel_factors}. Values below unity correspond to Zeno
    suppression, values above unity to anti-Zeno enhancement, and the ratio
    approaches unity in the long-time Markovian limit.
}
\label{fig:model_schematic}
\end{figure*}

The two reservoirs are held at temperatures \(T_h\) (hot) and \(T_c\) (cold). The TLS couples longitudinally
to the piston and transversely to both reservoirs, while the reservoirs couple neither to the piston nor to each other. The total Hamiltonian is
\begin{equation}
\HTot = \HS+\HP+\HSP+\HB+\HSB.
\label{eq:total_hamiltonian}
\end{equation}
Here \(\HS\) is the bare TLS Hamiltonian, \(\HP\) the piston Hamiltonian,
\(\HSP\) their mutual coupling, \(\HB\) the reservoir Hamiltonian, and
\(\HSB\) the TLS--reservoir interaction. The bare TLS and piston Hamiltonians
are
\begin{equation}
\HS = \frac{\hbar\omz}{2}\sigz, \qquad
\HP = \hbar\nup a^\dagger a,
\label{eq:bare_TLS_piston_hamiltonian}
\end{equation}
where \(\omz\) is the TLS transition frequency, \(\nup\) the piston
frequency, and \(a\) and \(a^\dagger\) the piston annihilation and creation
operators, with \([a,a^\dagger]=1\). We denote the TLS
ground and excited states by \(\ket{g}\) and \(\ket{e}\). In this basis
\(\sigz=\ket{e}\bra{e}-\ket{g}\bra{g}\), while \(\sigm=\ket{g}\bra{e}\) and
\(\sigp=\ket{e}\bra{g}\) are the lowering and raising operators.

The TLS couples longitudinally to the piston~\cite{GelbwaserKurizkiPRE2014,Richer2016Longitudinal,Bera2021Longitudinal},
\begin{equation}
\HSP = \hbar g\sigz(a+a^\dagger),
\label{eq:TLS_piston_coupling}
\end{equation}
with \(g\) the TLS--piston coupling strength. Because \(\HSP\) is proportional
to \(\sigz\), it commutes with \(\HS\) and does not induce TLS transitions by itself. Its
effect is to displace the oscillator equilibrium by a TLS-state-dependent
amount. The reservoirs are independent collections of bosonic modes,
\begin{equation}
\HB = \sum_{j=h,c} \sum_k \hbar\omega_{jk}b_{jk}^\dagger b_{jk},
\label{eq:bath_hamiltonian}
\end{equation}
where \(b_{jk}\) annihilates an excitation of frequency \(\omega_{jk}\) in
reservoir \(j\), and each reservoir starts in a thermal state \(\rho_{B_j}\) at
inverse temperature \(\beta_j=(k_B T_j)^{-1}\). The TLS couples transversely to
both reservoirs,
\begin{equation}
\HSB = \sum_{j=h,c} \sigx\otimes B_j, \qquad
B_j = \sum_k \hbar\lambda_{jk} \left( b_{jk}+b_{jk}^\dagger \right),
\label{eq:TLS_bath_coupling}
\end{equation}
with \(\lambda_{jk}\) the coupling strength to mode \(k\) of reservoir \(j\).
Since \(\sigx=\sigp+\sigm\), this coupling drives TLS transitions and thereby
lets the reservoirs exchange energy with the TLS--piston system. The model is summarized in Fig.~\ref{fig:model_schematic}.

\subsection{Finite-time sideband master equation}
\label{subsec:finite_time_sideband_master_equation}

Starting from the microscopic Hamiltonian in Eq.~\eqref{eq:total_hamiltonian}, we derive the finite-time reduced dynamics of
the coupled TLS--piston system. We first diagonalize the TLS--piston Hamiltonian and resolve the dressed
reservoir interaction into carrier and first-sideband transition channels.
We then obtain their finite-time dissipative dynamics using a second-order time-convolutionless treatment of the TLS--reservoir coupling.
The physical construction and resulting master equation are presented here, while the complete microscopic derivation is given
in Appendix~\ref{app:finite_time_master_equation}.

The longitudinal interaction in Eq.~\eqref{eq:TLS_piston_coupling} displaces the piston by an amount that depends on the TLS state. Since \(\HSP\) does not
commute with \(\HP\), the bare TLS--piston product states are not eigenstates of the coupled system. We therefore diagonalize the TLS--piston Hamiltonian using
the polaron transformation
\begin{equation}
U_p = \exp\!\left[ \zeta\sigz(a^\dagger-a) \right],\qquad
\zeta=\frac{g}{\nup}.
\label{eq:polaron_transformation_main}
\end{equation}
We use the dressed-operator convention \(\widetilde{O}=U_p^\dagger O U_p\), which gives
\begin{equation}
\tilde{a} = U_p^\dagger a U_p = a+\zeta\sigz, \qquad
\tilde{a}^\dagger = U_p^\dagger a^\dagger U_p = a^\dagger+\zeta\sigz.
\label{eq:polaron_ladder_operators_main}
\end{equation}
Since \(\sigz\) commutes with \(U_p\), \(\tilde{\sigma}_z=\sigz\). In terms of the dressed operators, the coupled
TLS--piston Hamiltonian becomes
\begin{equation}
\HS+\HP+\HSP = \frac{\hbar\omz}{2}\tilde{\sigma}_z + \hbar\nup\tilde{a}^\dagger\tilde{a} - \frac{\hbar g^2}{\nup} \equiv \widetilde{H}_{\rm sys}.
\label{eq:polaron_system_hamiltonian_main}
\end{equation}
The last term is a constant polaron shift and does not affect the reduced dynamics. Throughout the paper, dressed quantities are marked with a tilde.

We next express the TLS--reservoir interaction in this dressed representation.
Defining \(\tilde{X}=\tilde{a}^\dagger-\tilde{a}=a^\dagger-a\), the inverse relations between the bare and dressed TLS transition operators are
\begin{equation}
\sigp = \tilde{\sigma}_+e^{2\zeta\tilde{X}}, \qquad
\sigm = \tilde{\sigma}_-e^{-2\zeta\tilde{X}}.
\label{eq:polaron_sigma_pm_main}
\end{equation}
Equation~\eqref{eq:TLS_bath_coupling} therefore becomes
\begin{equation}
\widetilde{H}_{SB} = \sum_{j=h,c} \left(\tilde{\sigma}_+e^{2\zeta\tilde{X}} + \tilde{\sigma}_-e^{-2\zeta\tilde{X}} \right)\otimes B_j.
\label{eq:polaron_transformed_HSB_main}
\end{equation}
The displacement operators in Eq.~\eqref{eq:polaron_transformed_HSB_main} allow a reservoir-induced TLS
transition to exchange quanta with the piston. For \(\zeta\ll1\), the leading processes are obtained by expanding these operators to first order in \(\zeta\),
\begin{equation}
e^{\pm2\zeta\tilde{X}} \simeq 1\pm2\zeta\tilde{X}.
\label{eq:first_sideband_expansion_main}
\end{equation}
This approximation is distinct from the weak TLS--reservoir coupling used in the master-equation derivation. It truncates the transformed interaction to
the carrier and first pair of sidebands. Since the relevant matrix elements increase with the piston occupation, the truncation requires
\begin{equation}
2\zeta \sqrt{ \left\langle \tilde{a}^\dagger\tilde{a} \right\rangle_s+1
}\ll1
\label{eq:first_sideband_validity_condition_main}
\end{equation}
throughout the evolution considered. Applying the first-order expansion in Eq.~\eqref{eq:first_sideband_expansion_main} to the dressed interaction in
Eq.~\eqref{eq:polaron_transformed_HSB_main}, we decompose the resulting system operators into eigenoperators of
\(\widetilde{H}_{\rm sys}\) in Eq.~\eqref{eq:polaron_system_hamiltonian_main}. This decomposition yields the
carrier and first-sideband transition frequencies. For \(\omz>\nup\), the positive frequencies are
\begin{equation}
\Omega_{\rm sb} = \{\omz,\omm,\omp\}, \qquad \omm=\omz-\nup, \qquad \omp=\omz+\nup,
\label{eq:sideband_frequency_set_main}
\end{equation}
with corresponding positive-frequency transition operators
\begin{equation}
\begin{aligned}
\widetilde{A}(\omz) &= \tilde{\sigma}_-, \qquad \widetilde{A}(\omm) = -2\zeta\,\tilde{\sigma}_-\tilde{a}^\dagger, \\
\widetilde{A}(\omp)
&= 2\zeta\,\tilde{\sigma}_-\tilde{a}.
\end{aligned}
\label{eq:sideband_transition_operators_main}
\end{equation}
The carrier operator \(\widetilde{A}(\omz)\) relaxes the TLS without changing
the piston occupation. During TLS relaxation, the lower-sideband operator
\(\widetilde{A}(\omm)\) creates one piston quantum, whereas the upper-sideband
operator \(\widetilde{A}(\omp)\) annihilates one.

The reservoirs remain continuously coupled during the evolution. We use
\(s\) for the elapsed time and \(\tau_c\) for the selected endpoint at which
finite-time or coupling-averaged quantities are evaluated, with
\(0\le s\le\tau_c\). No switching operation is implied at \(s=\tau_c\).
Finite-memory corrections arise because the reservoir response entering the
time-local generator is accumulated only over the elapsed interval. The
corresponding correlation functions and finite-time response coefficients are
derived explicitly in Appendix~\ref{app:finite_time_master_equation}.

We obtain the reduced dynamics using a second-order time-convolutionless
(TCL2) expansion in the TLS--reservoir coupling
~\cite{BreuerPetruccione2002}. The TCL2 treatment assumes weak
TLS--reservoir coupling, an initially factorized TLS--piston--reservoir state,
and two independent reservoirs in stationary thermal states. We further apply
the secular approximation between distinct  transition frequencies,
which requires the carrier and sidebands to remain spectrally resolved.
Independently, the carrier-plus-first-sideband truncation requires
Eq.~\eqref{eq:first_sideband_validity_condition_main} to hold throughout the
evolution. Under these assumptions, the reduced TLS--piston density operator
obeys
\begin{align}
\frac{d\widetilde{\rho}(s)}{ds} ={}& -\frac{i}{\hbar} \left[ \widetilde{H}_{\rm sys} + \widetilde{H}_{\rm LS}(s), \widetilde{\rho}(s) \right] \nonumber\\
&+ \sum_{j=h,c} \sum_{\omega\in\Omega_{\rm sb}} \mathcal{L}_{j,\omega}(s)\widetilde{\rho}(s).
\label{eq:finite_time_sideband_master_equation}
\end{align}
The finite-time Lamb-shift Hamiltonian
\(\widetilde{H}_{\rm LS}(s)\) is given in
Eq.~\eqref{eq:app_lamb_shift}. For each positive transition frequency
\(\omega\in\Omega_{\rm sb}\), the dissipative contribution from reservoir
\(j\) is
\begin{align}
\mathcal{L}_{j,\omega}(s)\widetilde{\rho}
={}&
\gamma_j(\omega,s)
\mathcal{D}[\widetilde{A}(\omega)]\widetilde{\rho}
+
\gamma_j(-\omega,s)
\mathcal{D}[\widetilde{A}^{\dagger}(\omega)]
\widetilde{\rho}.
\label{eq:finite_time_sideband_liouvillian}
\end{align}
Here \(\gamma_j(\omega,s)\) governs the downward transition
\(\widetilde{A}(\omega)\), while \(\gamma_j(-\omega,s)\) governs the
corresponding upward transition
\(\widetilde{A}^{\dagger}(\omega)\). The Lindblad dissipator is
\begin{equation}
\mathcal{D}[L]\rho
=
L\rho L^\dagger
-
\frac{1}{2}
\left\{
L^\dagger L,\rho
\right\},
\label{eq:lindblad_dissipator_definition}
\end{equation}
and the finite-time rate is
\begin{equation}
\gamma_j(\omega,s)
=
2
\int_{-\infty}^{\infty}
d\xi\,
G_j(\xi)
\frac{
\sin[(\omega-\xi)s]
}{
\omega-\xi
}.
\label{eq:finite_time_rate_main}
\end{equation}
Here \(G_j(\xi)\) is the two-sided thermal response spectrum of reservoir
\(j\), defined microscopically in Eq.~\eqref{eq:app_two_sided_spectrum}.
The sinc-shaped kernel in Eq.~\eqref{eq:finite_time_rate_main} is centered at the transition frequency
\(\omega\) and has a characteristic width of order \(1/s\). The finite-time
rate therefore measures the spectral overlap between the system transition
and the reservoir response over the elapsed evolution time.

Because the sinc kernel in Eq.~\eqref{eq:finite_time_rate_main} has oscillatory side lobes,
the finite-time TCL2 coefficients are not guaranteed to remain nonnegative for arbitrary spectra and local times. We therefore restrict the operating regimes considered throughout the work to intervals in which all retained coefficients
\(\gamma_j(\pm\omega,s)\) remain nonnegative. Within such intervals, a
finite-time rate above its long-time golden-rule value corresponds to
anti-Zeno enhancement, whereas a smaller rate corresponds to Zeno
suppression. For \(s\gg\tau_B\), the kernel approaches the delta-function
limit derived in Eq.~\eqref{eq:app_markov_delta_limit}, giving
\(\gamma_j(\omega,s)\to2\pi G_j(\omega)\). Correspondingly, an evaluation
endpoint satisfying \(\tau_c\gg\tau_B\) lies in the long-time Markovian
regime.

\section{Quantum Heat Engine}
\label{sec:quantum_heat_engine}

We now specialize the finite-time master
equation~\eqref{eq:finite_time_sideband_master_equation} to the heat-engine
regime. We first establish the relevant time scales and spectral selection, then derive
an effective piston amplifier and its gain condition. We subsequently identify
ergotropy as the useful work output and quantify its anti-Zeno enhancement
relative to the Markovian reference.

\subsection{Evaluation time and operating time scales}
\label{subsec:engine_protocol_timescales}

We use the finite-time convention established in
Sec.~\ref{subsec:finite_time_sideband_master_equation}. For the engine
analysis, the carrier and first sidebands in Eq.~\eqref{eq:finite_time_sideband_master_equation}
must remain spectrally resolved while finite-memory effects are still appreciable. For the retained signed transition frequencies
\(\{\pm\omm,\pm\omz,\pm\omp\}\), the smallest separation is \(\Delta_{\rm sb}=\min\!\left\{\nup,\,2\omm\right\}\).
The relevant operating regime is therefore
\begin{equation}
\Delta_{\rm sb}^{-1}
\ll
\tau_c
\lesssim
\tau_B
\ll
\tau_{\rm rel},
\label{eq:engine_timescale_hierarchy}
\end{equation}
where \(\tau_B\) is the dominant reservoir correlation time and
\(\tau_{\rm rel}\sim\gamma_{\max}^{-1}\) is the relaxation time set by the
largest relevant retained dissipative rate.

Each inequality in Eq.~\eqref{eq:engine_timescale_hierarchy} has a distinct role. The condition
\(\Delta_{\rm sb}^{-1}\ll\tau_c\) ensures consistency with the secular
approximation used in the derivation of the master equation~\eqref{eq:finite_time_sideband_master_equation}.
The condition \(\tau_c\lesssim\tau_B\) keeps finite-memory corrections
appreciable, while their suppression or enhancement is determined by the
finite-time spectral overlap in Eq.~\eqref{eq:finite_time_rate_main}.
Finally, \(\tau_B\ll\tau_{\rm rel}\) ensures that reservoir
correlations decay before the reduced TLS--piston state changes appreciably,
as required by the weak-coupling Born and TCL2 treatment. The operating window
considered below therefore lies between the unresolved short-time regime
\(\Delta_{\rm sb}\tau_c\lesssim1\) and the long-time Markovian regime
\(\tau_c\gg\tau_B\).

\subsection{Engine regime and sideband selection}
\label{subsec:engine_sideband_selection}

The reservoir spectra determine which of the carrier and first-sideband
transitions identified in
Eqs.~\eqref{eq:sideband_frequency_set_main} and
\eqref{eq:sideband_transition_operators_main} contribute appreciably through
the finite-time spectral overlap in Eq.~\eqref{eq:finite_time_rate_main}.
The spectral design therefore determines the dominant channels, while the
relative strengths of the forward and reverse processes determine the
preferred cycle direction. Consequently, the same dominant-channel
configuration can support either engine or refrigerator
operation~\cite{GelbwaserPRE2013,GelbwaserKurizkiPRE2014}.

For the configuration considered here, the hot reservoir predominantly drives
the carrier transition, while the cold reservoir predominantly drives the
lower sideband,
\begin{equation}
G_h(\omz)\ \text{dominant},
\qquad
G_c(\omm)\ \text{dominant},
\label{eq:engine_spectral_selection}
\end{equation}
with the remaining positive-frequency overlaps spectrally suppressed,
\begin{align}
G_h(\omm),\;G_h(\omp)&\ll G_h(\omz),
\nonumber\\
G_c(\omz),\;G_c(\omp)&\ll G_c(\omm).
\label{eq:engine_suppressed_channels}
\end{align}
Because the finite-time rates in
Eq.~\eqref{eq:finite_time_rate_main} sample a finite spectral interval,
pointwise suppression alone is insufficient. The retained-channel
approximation in Eq.~\eqref{eq:engine_suppressed_channels} therefore requires
the unwanted finite-time rates to remain small relative to the hot-carrier
and cold-lower-sideband rates throughout the operating window, as verified in
Sec.~\ref{subsec:AZD_piston_ergotropy}.

With this channel selection, the elementary forward cycle is
\begin{equation}
\ket{g,n}
\xrightarrow{\;h,\omz\;}
\ket{e,n}
\xrightarrow{\;c,\omm\;}
\ket{g,n+1},
\label{eq:engine_forward_cycle}
\end{equation}
where \(\ket{g,n}\) and \(\ket{e,n}\) are dressed joint eigenstates of
\(\widetilde{H}_{\rm sys}\) in
Eq.~\eqref{eq:polaron_system_hamiltonian_main}, with \(n\) the dressed piston
occupation. In the first step, the TLS absorbs energy \(\hbar\omz\) from the
hot reservoir at the carrier frequency. It then relaxes through the cold
lower-sideband channel, transferring one quantum \(\hbar\nup\) to the piston
and energy \(\hbar\omm=\hbar\omz-\hbar\nup\) to the cold reservoir.

Neglecting the finite-time Lamb shift and moving to the interaction picture
generated by \(\widetilde{H}_{\rm sys}\), the spectral conditions in
Eqs.~\eqref{eq:engine_spectral_selection} and
\eqref{eq:engine_suppressed_channels} reduce the general master equation
\eqref{eq:finite_time_sideband_master_equation} to
\begin{align}
\dot{\widetilde{\rho}}(s)
= {}&
r_h^\downarrow(s)\D[\tilde{\sigma}_-]\widetilde{\rho}(s)
+
r_h^\uparrow(s)\D[\tilde{\sigma}_+]\widetilde{\rho}(s)
\nonumber\\
&+
r_c^\downarrow(s)
\D[\tilde{\sigma}_-\tilde{a}^\dagger]\widetilde{\rho}(s)
+
r_c^\uparrow(s)
\D[\tilde{\sigma}_+\tilde{a}]\widetilde{\rho}(s).
\label{eq:engine_master_equation_compact}
\end{align}
The hot-carrier rates are
\begin{alignat}{2}
r_h^\downarrow(s) &= \gamma_h(\omz,s),
\qquad&
r_h^\uparrow(s) &= \gamma_h(-\omz,s),
\label{eq:engine_hot_rates}
\end{alignat}
while the cold lower-sideband rates are
\begin{alignat}{2}
r_c^\downarrow(s) &= 4\zeta^2\gamma_c(\omm,s),
\qquad&
r_c^\uparrow(s) &= 4\zeta^2\gamma_c(-\omm,s).
\label{eq:engine_cold_rates}
\end{alignat}
The factor \(4\zeta^2\) in Eq.~\eqref{eq:engine_cold_rates} originates from
the squared first-order sideband amplitudes in
Eq.~\eqref{eq:sideband_transition_operators_main}.
The forward cycle in Eq.~\eqref{eq:engine_forward_cycle} combines hot-carrier
excitation with cold-sideband relaxation, whereas the reverse cycle combines
cold-sideband excitation with hot-carrier relaxation. For a fixed pair of
adjacent piston levels, the sideband matrix-element factor is the same in both
directions and therefore cancels when the two cycles are compared. The
elementary cycle is thus locally biased in the engine direction when
\(
r_h^\uparrow(s)r_c^\downarrow(s)
>
r_h^\downarrow(s)r_c^\uparrow(s)
\).
This rate-only cycle-bias condition should be distinguished from the
population-dependent condition for net piston amplification derived in
Sec.~\ref{subsec:piston_amplification_ergotropy}.

\subsection{Piston amplification and ergotropy storage}
\label{subsec:piston_amplification_ergotropy}

Here we reduce the retained TLS--piston dynamics in
Eq.~\eqref{eq:engine_master_equation_compact} to an effective piston amplifier
and derive its gain condition. We then distinguish piston energy from
extractable work through the ergotropy. For an initially coherent piston, the
ergotropy can be obtained directly from the coherent displacement, while
other initial piston states are treated in
Appendix~\ref{app:piston_state_ergotropy}.

The hot carrier and cold lower sideband in Eq.~\eqref{eq:engine_master_equation_compact}
play distinct dynamical roles. The hot carrier predominantly sets the TLS populations. 
The cold lower sideband converts the resulting population bias into piston amplification or damping.
The cold-sideband rates in Eq.~\eqref{eq:engine_cold_rates} carry the
prefactor \(4\zeta^2\). Since \(\zeta\ll1\), this
suppresses the sideband coupling relative to the carrier scale. For the parameter regime considered in
Sec.~\ref{subsec:AZD_piston_ergotropy}, the resulting cold-sideband piston
dynamics remain slower than the hot-carrier relaxation, so the sideband
backaction on the TLS is weak. This time-scale separation motivates the
product-state approximation
\begin{equation}
\widetilde{\rho}(s)
\simeq
\rho_{\rm TLS}(s)\otimes\rho_{P}(s).
\label{eq:piston_level_reduction}
\end{equation}
Here
\(\rho_{\rm TLS}(s)=\Tr_{P}[\widetilde{\rho}(s)]\) and
\(\rho_{P}(s)=\Tr_{\rm TLS}[\widetilde{\rho}(s)]\). The instantaneous TLS
populations are
\begin{equation}
p_{e}(s)
=
\langle e|\rho_{\rm TLS}(s)|e\rangle,
\qquad
p_{g}(s)=1-p_{e}(s).
\label{eq:TLS_populations_engine}
\end{equation}
Within this reduction, the TLS enters the piston dynamics through these
populations. The accuracy of Eq.~\eqref{eq:piston_level_reduction} for the
parameter regime considered here is verified numerically in
Sec.~\ref{subsec:AZD_piston_ergotropy}.

Tracing over the TLS in the cold lower-sideband part of
Eq.~\eqref{eq:engine_master_equation_compact} gives a phase-insensitive
gain--loss equation for the
piston~\cite{Caves1982,GardinerZoller2004,GelbwaserEPL2013},
\begin{equation}
\dot{\rho}_{P}(s)
=
D_{P}(s)\,\mathcal{D}[\tilde{a}^{\dagger}]\rho_{P}(s)
+
\bigl[D_{P}(s)+\Gamma(s)\bigr]\,
\mathcal{D}[\tilde{a}]\rho_{P}(s).
\label{eq:effective_piston_amplifier}
\end{equation}
The piston-excitation coefficient is
\begin{equation}
D_{P}(s)
=
r_{c}^{\downarrow}(s)\,p_{e}(s),
\label{eq:piston_diffusion_coefficient}
\end{equation}
and describes the forward cold-sideband process
\(\ket{e,n}\to\ket{g,n+1}\), which creates one piston quantum. The net drift
coefficient is
\begin{equation}
\Gamma(s)
=
r_{c}^{\uparrow}(s)\,p_{g}(s)
-
r_{c}^{\downarrow}(s)\,p_{e}(s).
\label{eq:piston_drift_coefficient}
\end{equation}
Its two terms describe piston removal and creation, respectively, so coherent
amplification requires \(\Gamma(s)<0\). Using
Eq.~\eqref{eq:piston_drift_coefficient} together with the cold-sideband rates
in Eq.~\eqref{eq:engine_cold_rates}, this condition becomes
\begin{equation}
\frac{p_{e}(s)}{p_{g}(s)}
>
\frac{\gamma_{c}(-\omm,s)}{\gamma_{c}(\omm,s)}.
\label{eq:piston_gain_condition_finite_time_rates}
\end{equation}
Thus, the TLS excited-to-ground population ratio must exceed the
reverse-to-forward cold-sideband rate ratio. Unlike the cycle-bias condition
in Sec.~\ref{subsec:engine_sideband_selection},
Eq.~\eqref{eq:piston_gain_condition_finite_time_rates} explicitly includes
the instantaneous TLS populations and therefore determines the actual piston
gain.

The coherent amplitude and mean piston occupation respond differently to the
gain--loss dynamics governed by Eq.~\eqref{eq:effective_piston_amplifier}. Accordingly, the coherent amplitude
\(
\alpha(s)\equiv\langle\tilde{a}\rangle_s
\)
obeys
\begin{equation}
\dot{\alpha}(s)
=
-\frac{\Gamma(s)}{2}\,\alpha(s),
\label{eq:piston_amplitude_equation}
\end{equation}
which integrates to
\begin{equation}
\alpha(s)
=
\alpha_{0}
\exp\!\left[
-\frac{1}{2}
\int_{0}^{s}ds'\,\Gamma(s')
\right].
\label{eq:piston_amplitude_solution}
\end{equation}
Following Eq.~\eqref{eq:effective_piston_amplifier}, the mean piston occupation,
\(
n_{P}(s)
=
\langle\tilde{a}^{\dagger}\tilde{a}\rangle_s
\),
instead satisfies
\begin{equation}
\dot{n}_{P}(s)
=
D_{P}(s)
-
\Gamma(s)\,n_{P}(s).
\label{eq:piston_number_dynamics}
\end{equation}
Thus, the coherent amplitude in Eq.~\eqref{eq:piston_amplitude_equation} is governed solely by the net drift
\(\Gamma(s)\), whereas the occupation in Eq.~\eqref{eq:piston_number_dynamics} also contains the additive excitation
term \(D_P(s)\). The piston energy can therefore increase through
phase-insensitive added noise without a corresponding increase in coherent
displacement.

Within the reduced dressed-mode description, the mean piston energy is
\begin{equation}
E_{P}(s)
=
\Tr\!\left[
\hbar\nup\,
\tilde{a}^{\dagger}\tilde{a}\,
\rho_{P}(s)
\right].
\label{eq:piston_total_energy}
\end{equation}
For an arbitrary piston state, the maximum work extractable by a cyclic
unitary transformation is quantified by the ergotropy~\cite{PuszWoronowicz1978,Lenard1978,Allahverdyan2004},
\begin{equation}
\WP[\rho_P(s)]
=
E_{P}(s)
-
E_{\rm pas}[\rho_P(s)].
\label{eq:piston_ergotropy_definition}
\end{equation}
Here \(E_{\rm pas}[\rho_P(s)]\) is the energy of the passive state obtained by
ordering the eigenvalues of \(\rho_P(s)\) decreasingly over increasing
oscillator energies. It is the minimum energy within the unitary orbit of
\(\rho_P(s)\)~\cite{PuszWoronowicz1978,Lenard1978,Allahverdyan2004}.
Consequently, an increase in piston energy does not necessarily imply an
increase in extractable work~\cite{GelbwaserKurizkiPRE2014}.

We now consider the initially coherent piston state
\begin{equation}
\rho_{P}(0)
=
\ket{\alpha_{0}}\bra{\alpha_{0}}.
\label{eq:initial_coherent_piston_state}
\end{equation}
Under the phase-insensitive gain--loss dynamics of
Eq.~\eqref{eq:effective_piston_amplifier}, the coherent state evolves into a
displaced thermal state. Its centered thermal component is passive, so the
ergotropy is carried entirely by the coherent displacement. A detailed
derivation is given in Appendix~\ref{app:piston_state_ergotropy}, yielding
\begin{equation}
\WP^{\rm coh}(s)
=
\hbar\nup\,|\alpha(s)|^{2}.
\label{eq:coherent_state_ergotropy}
\end{equation}
Substituting Eq.~\eqref{eq:piston_amplitude_solution} into
Eq.~\eqref{eq:coherent_state_ergotropy} gives
\begin{equation}
\WP^{\rm coh}(s)
=
\hbar\nup\,|\alpha_{0}|^{2}
\exp\!\left[
-\int_{0}^{s}ds'\,\Gamma(s')
\right],
\label{eq:coherent_state_ergotropy_solution}
\end{equation}
and differentiation yields
\begin{equation}
\dot{\WP}\!^{\rm coh}(s)
=
-\Gamma(s)\,\WP^{\rm coh}(s).
\label{eq:ergotropy_growth_rate}
\end{equation}
Thus, the coherent piston ergotropy grows for \(\Gamma(s)<0\) and decays for
\(\Gamma(s)>0\). The additive excitation term \(D_P(s)\) appearing in
Eq.~\eqref{eq:piston_number_dynamics} does not enter independently in the
coherent-ergotropy dynamics. Instead, the added occupation contributes to the
passive thermal component of the displaced thermal state.

For general piston preparations, ergotropy cannot be inferred from
\(|\alpha(s)|^2\) alone and must instead be evaluated from
Eq.~\eqref{eq:piston_ergotropy_definition}. Consequently, enhanced
finite-time amplification does not necessarily imply enhanced ergotropy for
a noncoherent piston state. Appendix~\ref{app:piston_state_ergotropy}
examines this state dependence for coherent, squeezed, Fock, and thermal
preparations.

\subsection{Anti-Zeno enhancement of coherent piston ergotropy}
\label{subsec:AZD_piston_ergotropy}

To quantify the finite-time enhancement, we introduce the net gain coefficient
\(\Lambda(s)\equiv-\Gamma(s)\), for which coherent amplification corresponds
to \(\Lambda(s)>0\). Equation~\eqref{eq:ergotropy_growth_rate} then becomes
\begin{equation}
\dot{\WP}\!^{\rm coh}(s)
=
\Lambda(s)\,\WP\!^{\rm coh}(s).
\label{eq:AZD_ergotropy_growth_gain_form}
\end{equation}
We ask whether finite-time reservoir sampling can increase this gain relative
to its Markovian value and thereby generate more coherent piston ergotropy
over the same coupling interval \(\tau_c\).

\begin{figure*}[t]
    \centering
    \begin{minipage}[t]{0.325\textwidth}
        \centering
        \includegraphics[width=\linewidth]{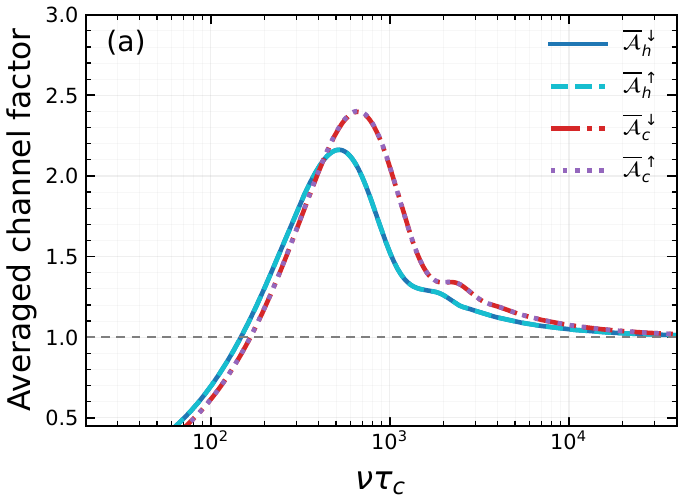}
    \end{minipage}
    \hfill
    \begin{minipage}[t]{0.325\textwidth}
        \centering
        \includegraphics[width=\linewidth]{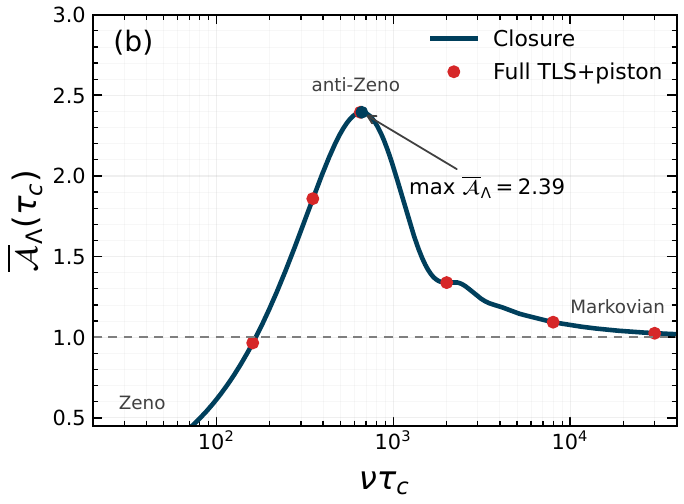}
    \end{minipage}
    \hfill
    \begin{minipage}[t]{0.325\textwidth}
        \centering
        \includegraphics[width=\linewidth]{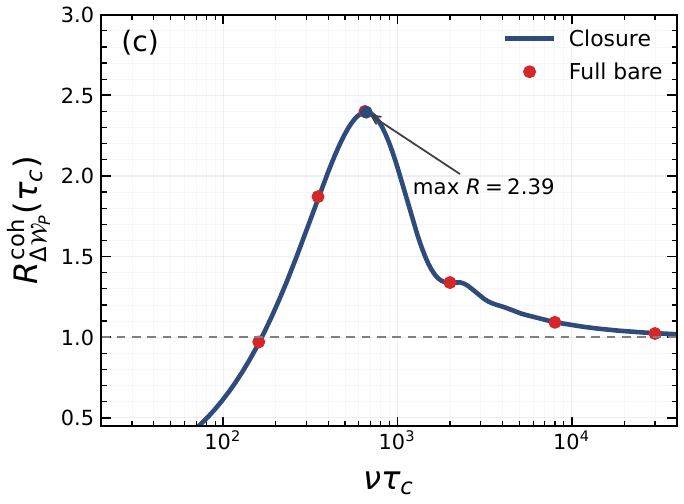}
    \end{minipage}
    \caption{
    Anti-Zeno enhancement of coherent piston ergotropy in the heat-engine
    regime. The horizontal dashed line marks the Markovian reference value
    of unity, and the horizontal axis is the dimensionless coupling time
    \(\nup\tau_c\) on a logarithmic scale.
    (a) Coupling-averaged retained-channel factors
    \(\overline{\mathcal{A}}_{h}^{\downarrow,\uparrow}\) and
    \(\overline{\mathcal{A}}_{c}^{\downarrow,\uparrow}\), defined in
    Eq.~\eqref{eq:AZD_channel_factors}.
    (b) Coupling-averaged net gain factor
    \(\overline{\mathcal{A}}_\Lambda(\tau_c)\), defined in
    Eq.~\eqref{eq:AZD_gain_enhancement_factor_Lambda}, with
    \(\max\overline{\mathcal{A}}_\Lambda\simeq2.39\) near
    \(\nup\tau_c\simeq662\).
    (c) Generated coherent ergotropy ratio
    \(R_{\Delta\WP}^{\rm coh}(\tau_c)\), defined in
    Eq.~\eqref{eq:AZD_fixed_time_ergotropy_ratio_compact}, with maximum
    \(\simeq2.39\).
    The markers in panels~(b) and (c) are obtained by numerically solving the
    full joint TLS--piston master equation
    \eqref{eq:engine_master_equation_compact}. In panel~(c), the markers use
    the bare-frame piston ergotropy obtained by undoing the polaron
    transformation and applying the passive-state construction in
    Eq.~\eqref{eq:piston_ergotropy_definition}.
    Parameters are given in dimensionless units with
    \(\hbar=k_B=\nup=1\):
    \(\omz=3\nup\), \(\omm=2\nup\), \(\zeta=0.095\),
    \(|\alpha_0|=1\), \(\beta_h\omz=0.35\), and
    \(\beta_c\omm=1.50\).
    The filtered Lorentzian spectra in
    Eqs.~\eqref{eq:AZD_filtered_hot_spectrum} and
    \eqref{eq:AZD_filtered_cold_spectrum} use
    \(G_{0h}=G_{0c}=10^{-5}\),
    \(\Gamma_h=1.60\times10^{-3}\nup\),
    \(\Gamma_c=1.10\times10^{-3}\nup\),
    \(\delta_h=5.00\times10^{-3}\nup\),
    \(\delta_c=3.80\times10^{-3}\nup\),
    \(\Delta_{w,h}=8.50\times10^{-3}\nup\), and
    \(\Delta_{w,c}=7.50\times10^{-3}\nup\).
    The joint-solve piston cutoff is \(N_P=32\).
    }
    \label{fig:AZD_piston_ergotropy}
\end{figure*}

To obtain a closed analytical expression for the gain, we use the dynamical
separation established in
Sec.~\ref{subsec:piston_amplification_ergotropy}. In Eq.~\eqref{eq:engine_master_equation_compact},
the hot carrier provides the dominant contribution to the TLS population dynamics, while the cold-sideband
backaction remains weak in the parameter regime considered in Fig.~\ref{fig:AZD_piston_ergotropy}. We therefore
approximate the TLS population bias by the local stationary bias associated
with the hot-carrier channel. Setting the local hot-carrier population flow to
zero gives
\begin{align}
p_e^h(s)
&=
\frac{\gamma_h(-\omz,s)}
{\gamma_h(\omz,s)+\gamma_h(-\omz,s)},
\nonumber\\[4pt]
p_g^h(s)
&=
\frac{\gamma_h(\omz,s)}
{\gamma_h(\omz,s)+\gamma_h(-\omz,s)}.
\label{eq:AZD_hot_population_closure}
\end{align}
Equation~\eqref{eq:AZD_hot_population_closure} is understood for \(s>0\),
with its value at \(s=0\) defined by the \(s\to0^+\) limit. It is used only as
an analytical population closure and does not assume that the TLS dynamically
equilibrates within the finite coupling interval.

Within this closure, substitution into
Eq.~\eqref{eq:piston_drift_coefficient}, together with the cold-sideband rates
in Eq.~\eqref{eq:engine_cold_rates}, gives the finite-time gain
\begin{equation}
\Lambda_{\rm FT}(s)
=
r_c^\downarrow(s)\,p_e^h(s)
-
r_c^\uparrow(s)\,p_g^h(s).
\label{eq:AZD_positive_gain_direct}
\end{equation}
Its accumulation over a coupling interval \(\tau_c\) is
\begin{equation}
K_\Lambda(\tau_c)
=
\int_0^{\tau_c} ds\,
\Lambda_{\rm FT}(s),
\label{eq:AZD_integrated_gain_KLambda}
\end{equation}
with corresponding coupling-averaged gain
\begin{equation}
\overline{\Lambda}_{\rm FT}(\tau_c)
=
\frac{K_\Lambda(\tau_c)}{\tau_c}.
\label{eq:AZD_contact_averaged_gain}
\end{equation}
The Markovian reference is obtained by replacing the finite-time rates in both
the hot-population closure and the cold-sideband dynamics by their long-time
golden-rule values,
\begin{equation}
\Lambda_{\rm M}
=
r_{c,{\rm M}}^\downarrow\,p_e^{h,{\rm M}}
-
r_{c,{\rm M}}^\uparrow\,p_g^{h,{\rm M}}.
\label{eq:AZD_markovian_positive_gain}
\end{equation}
For \(\tau_c\gg\tau_B\),
Eq.~\eqref{eq:finite_time_rate_main} approaches its Markovian limit, so
\(\overline{\Lambda}_{\rm FT}(\tau_c)\to\Lambda_{\rm M}\). For the benchmark
considered here, \(\Lambda_{\rm M}>0\). We quantify the finite-time
modification of the net gain by
\begin{equation}
\overline{\mathcal{A}}_\Lambda(\tau_c)
=
\frac{\overline{\Lambda}_{\rm FT}(\tau_c)}
{\Lambda_{\rm M}}.
\label{eq:AZD_gain_enhancement_factor_Lambda}
\end{equation}
Within the positive-gain engine regime,
\(0<\overline{\mathcal{A}}_\Lambda<1\) corresponds to Zeno suppression of the
net gain, while \(\overline{\mathcal{A}}_\Lambda>1\) corresponds to anti-Zeno
enhancement.

The finite-time gain in Eq.~\eqref{eq:AZD_positive_gain_direct} determines the coherent ergotropy generated over the
same coupling interval. Defining
\(
\Delta\WP^{\rm coh}(\tau_c)
=
\WP^{\rm coh}(\tau_c)-\WP^{\rm coh}(0)
\),
Eq.~\eqref{eq:coherent_state_ergotropy_solution} gives, for identical initial
coherent piston states in the finite-time and Markovian evolutions,
\begin{equation}
R_{\Delta\WP}^{\rm coh}(\tau_c)
=
\frac{
\exp\!\left[
\mu_\Lambda(\tau_c)\,
\overline{\mathcal{A}}_\Lambda(\tau_c)
\right]-1
}{
\exp[\mu_\Lambda(\tau_c)]-1
},
\label{eq:AZD_fixed_time_ergotropy_ratio_compact}
\end{equation}
where
\(\mu_\Lambda(\tau_c)\equiv\Lambda_{\rm M}\tau_c\) is the dimensionless
Markovian charging depth. Thus,
\(R_{\Delta\WP}^{\rm coh}>1\) means that the finite-time machine generates
more coherent piston ergotropy than the Markovian reference over the same
physical coupling interval.
{\color{blue}This enhancement also has a direct charging interpretation.
The average ergotropy charging power is
\(\overline{\mathcal{P}}_{\rm erg}^{\rm coh}(\tau_c)
=\Delta\WP^{\rm coh}(\tau_c)/\tau_c\), where subtracting the initial
ergotropy isolates the work resource added during operation. At the
same coupling time, \(R_{\Delta\WP}^{\rm coh}\) is therefore also the
ratio of the finite-time and Markovian average ergotropy charging powers.}
In the weak-charging benchmark considered in
Fig.~\ref{fig:AZD_piston_ergotropy},
\(\mu_\Lambda(\tau_c)\ll1\) and \(K_\Lambda(\tau_c)\ll1\), so
Eq.~\eqref{eq:AZD_fixed_time_ergotropy_ratio_compact} becomes
\begin{equation}
R_{\Delta\WP}^{\rm coh}(\tau_c)
\simeq
\overline{\mathcal{A}}_\Lambda(\tau_c).
\label{eq:AZD_weak_charging_ergotropy_gain_relation}
\end{equation}

For comparison with the net gain, we also characterize the individual
retained channels. Averaging the finite-time rate in
Eq.~\eqref{eq:finite_time_rate_main} over the coupling interval gives
\begin{equation}
\overline{\gamma}_j(\omega,\tau_c)
=
\frac{1}{\tau_c}
\int_0^{\tau_c} ds\,
\gamma_j(\omega,s).
\label{eq:AZD_contact_averaged_rate}
\end{equation}
Using the Markovian rate
\(
\gamma_j^{\rm M}(\omega)\equiv2\pi G_j(\omega)
\),
we define the retained-channel factors
\begin{align}
\overline{\mathcal{A}}_h^\downarrow(\tau_c)
&=
\frac{\overline{\gamma}_h(\omz,\tau_c)}
{\gamma_h^{\rm M}(\omz)},
&
\overline{\mathcal{A}}_h^\uparrow(\tau_c)
&=
\frac{\overline{\gamma}_h(-\omz,\tau_c)}
{\gamma_h^{\rm M}(-\omz)},
\nonumber\\[4pt]
\overline{\mathcal{A}}_c^\downarrow(\tau_c)
&=
\frac{\overline{\gamma}_c(\omm,\tau_c)}
{\gamma_c^{\rm M}(\omm)},
&
\overline{\mathcal{A}}_c^\uparrow(\tau_c)
&=
\frac{\overline{\gamma}_c(-\omm,\tau_c)}
{\gamma_c^{\rm M}(-\omm)}.
\label{eq:AZD_channel_factors}
\end{align}
These factors diagnose the finite-time modification of the individual
retained channels. The coherent ergotropy, however, is governed by the net
gain in Eq.~\eqref{eq:AZD_positive_gain_direct}.

To realize the retained-channel conditions in
Eqs.~\eqref{eq:engine_spectral_selection} and
\eqref{eq:engine_suppressed_channels}, we use filtered Lorentzian reservoir
spectra for the numerical benchmark. The positive-frequency hot response is \cite{MukherjeeCommunPhys2020}
\begin{equation}
G_h^{(+)}(\xi)
=
G_{0h}
\frac{\Gamma_h^2}
{
\left[\xi-(\omz+\delta_h)\right]^2+\Gamma_h^2
}
W_h(\xi),
\qquad
\xi>0,
\label{eq:AZD_filtered_hot_spectrum}
\end{equation}
and the corresponding cold response is
\begin{equation}
G_c^{(+)}(\xi)
=
G_{0c}
\frac{\Gamma_c^2}
{
\left[\xi-(\omm-\delta_c)\right]^2+\Gamma_c^2
}
W_c(\xi),
\qquad
\xi>0.
\label{eq:AZD_filtered_cold_spectrum}
\end{equation}
Here \(\Gamma_h\) and \(\Gamma_c\) are the Lorentzian half widths at half
maximum, while \(\delta_h\) and \(\delta_c\) set the detunings of the spectral
maxima from the retained transitions. The filter windows are
\begin{align}
W_h(\xi)
&=
\Theta\!\left(\xi-\omz+\Delta_{w,h}\right)
\Theta\!\left(\omz+\Delta_{w,h}-\xi\right),
\label{eq:AZD_hot_window}
\\
W_c(\xi)
&=
\Theta\!\left(\xi-\omm+\Delta_{w,c}\right)
\Theta\!\left(\omm+\Delta_{w,c}-\xi\right).
\label{eq:AZD_cold_window}
\end{align}
The negative-frequency branches follow from the thermal detailed-balance
relation in Eq.~\eqref{eq:app_detailed_balance}. The windows restrict the
dominant hot and cold responses to the carrier and lower-sideband regions,
respectively, while the small detunings displace the Lorentzian maxima from
the corresponding transition frequencies. At intermediate coupling times,
the finite-width kernel in Eq.~\eqref{eq:finite_time_rate_main} can therefore
sample the nearby spectral peaks more strongly than the long-time
delta-function limit.

This construction produces the Zeno, anti-Zeno, and Markovian regimes shown
in Fig.~\ref{fig:AZD_piston_ergotropy}. At short coupling times, the
finite-time rates are suppressed relative to their Markovian values. At
intermediate times, the broadened kernel in Eq.~\eqref{eq:finite_time_rate_main}
overlaps the detuned spectral peaks more strongly, producing anti-Zeno enhancement. For
\(\tau_c\gg\tau_B\), the delta-function limit is recovered, and the normalized
rates approach unity. Figure~\ref{fig:AZD_piston_ergotropy}(a) shows the four retained-channel
factors in Eq.~\eqref{eq:AZD_channel_factors}. Their quantitative differences
reflect the distinct reservoir linewidths and detunings, while the upward and
downward factors within each reservoir remain close because the thermal factor
varies only weakly across the narrow spectral support. All four exhibit the
same qualitative Zeno--anti-Zeno--Markovian crossover.

Figure~\ref{fig:AZD_piston_ergotropy}(b) shows the corresponding net gain
factor \(\overline{\mathcal{A}}_\Lambda(\tau_c)\) in Eq.~\eqref{eq:AZD_gain_enhancement_factor_Lambda}.
It rises above unity after the Zeno-suppressed region, reaches
\(\max\overline{\mathcal{A}}_\Lambda\simeq2.39\) at \(\nup\tau_c\simeq662\),
and approaches unity in the long-time Markovian regime. The numerical solutions of the full retained-channel TLS--piston master
equation in Eq.~\eqref{eq:engine_master_equation_compact} agree with the
analytical prediction for \(\overline{\mathcal{A}}_\Lambda(\tau_c)\) in
Eq.~\eqref{eq:AZD_gain_enhancement_factor_Lambda}. The agreement is better
than \(0.7\%\) over the displayed marker set, and the two essentially
coincide near the maximum. This agreement supports the
population closure and product-state reduction for the benchmark considered.

Figure~\ref{fig:AZD_piston_ergotropy}(c) shows the generated coherent
ergotropy ratio defined in
Eq.~\eqref{eq:AZD_fixed_time_ergotropy_ratio_compact}.
Its maximum, \(R_{\Delta\WP}^{\rm coh}\simeq2.39\), means that the
finite-time machine generates approximately \(2.39\) times the coherent
piston ergotropy of the Markovian reference over the same coupling interval.
The corresponding increase in average ergotropy charging
power demonstrates enhanced useful charging of the piston under the
same initial preparation and reservoir spectra.
The close agreement with \(\overline{\mathcal{A}}_\Lambda(\tau_c)\) follows from the weak-charging
relation in Eq.~\eqref{eq:AZD_weak_charging_ergotropy_gain_relation}.
The numerical markers are obtained by solving the full retained-channel
TLS--piston master equation in Eq.~\eqref{eq:engine_master_equation_compact}. The resulting joint state
is then transformed back to the bare frame using
Eq.~\eqref{eq:polaron_transformation_main}, after which the reduced piston
ergotropy is evaluated from Eq.~\eqref{eq:piston_ergotropy_definition}.
Their agreement with the analytical curve shows that the dressed-mode
prediction remains quantitatively consistent with the bare-frame piston
ergotropy for this benchmark.

We also verify the validity of the quantitative anti-Zeno benchmark. The
displayed interval in Fig.~\ref{fig:AZD_piston_ergotropy} satisfies the resolved-sideband condition in Eq.~\eqref{eq:engine_timescale_hierarchy}, with \(\Delta_{\rm sb}\tau_c\ge20\). All retained instantaneous rates remain
nonnegative on the numerical grid. As a linewidth-based estimate of the
reservoir memory scale, we use \(\tau_{B,\Gamma}^{\max}\equiv \max(\Gamma_h^{-1},\Gamma_c^{-1})\), for which
\(\gamma_{\max}\tau_{B,\Gamma}^{\max}
\simeq1.56\times10^{-2}\).
The carrier-plus-first-sideband condition in
Eq.~\eqref{eq:first_sideband_validity_condition_main} is monitored using the
complete reduced-piston mean occupation, including the excitation contribution
\(D_P(s)\), and reaches \(\max_s[2\zeta(\langle\tilde{a}^\dagger\tilde{a}\rangle_s+1)^{1/2}]\simeq0.1\).
From the master equation~\eqref{eq:finite_time_sideband_master_equation}, with
the finite-time rates in Eq.~\eqref{eq:finite_time_rate_main}, we compare the
time-integrated absolute discarded and retained rate coefficients.
Relative to the retained rates in Eqs.~\eqref{eq:engine_hot_rates} and
\eqref{eq:engine_cold_rates}, the discarded-to-retained ratios are
approximately \(2.4\times10^{-4}\) for the hot reservoir and \(7.7\%\) for
the cold reservoir.

Thus, finite-time reservoir sampling enhances the net piston gain and the
coherent ergotropy generated over a fixed coupling interval.
This useful-work benefit depends on the piston preparation.
Appendix~\ref{app:piston_state_ergotropy} shows that, for the preparations
and parameters considered, finite-time dynamics enhances coherent-state
ergotropy accumulation but accelerates ergotropy loss for the squeezed
and Fock inputs. The thermal input remains passive under the same dynamics.
The enhancement of transition rates must therefore be
assessed together with the state-dependent balance between ergotropy
and passive energy. Because it modifies transition rates rather than channel energies, the
corresponding channel-energy efficiency remains fixed by the sideband
geometry, as discussed in Appendix~\ref{app:efficiency_cop}.

\section{Quantum Refrigerator}
\label{sec:quantum_refrigerator}

The finite-time master equation in
Eq.~\eqref{eq:finite_time_sideband_master_equation} also supports refrigerator
operation when the net cycle considered in
Sec.~\ref{sec:quantum_heat_engine} is reversed. In contrast to the engine cycle
in Eq.~\eqref{eq:engine_forward_cycle}, the piston now supplies the quantum
resource required to extract heat from the cold reservoir and transfer energy
to the hot reservoir. Refrigerator operation therefore consumes the finite
piston resource rather than charging it, as shown in Fig.~\ref{fig:model_schematic}.

We first identify the reverse cooling cycle and derive the corresponding finite
resource dynamics in Sec.~\ref{subsec:refrigerator_cycle_dynamics}. We then
show in Sec.~\ref{subsec:refrigerator_AZD_finite_resource} that instantaneous
finite-time reservoir sampling enhances both the cold current and the
accumulated extracted cold heat relative to the Markovian reference.

\subsection{Reverse cooling cycle and finite-resource dynamics}
\label{subsec:refrigerator_cycle_dynamics}

The refrigerator uses the same retained hot carrier and cold lower sideband
channels specified by Eqs.~\eqref{eq:engine_spectral_selection} and
\eqref{eq:engine_suppressed_channels}. The distinction from engine operation is
the direction of the net cycle. For \(n\geq1\), the elementary cooling cycle is
\begin{equation}
\ket{g,n}
\xrightarrow{\;c,\omm\;}
\ket{e,n-1}
\xrightarrow{\;h,\omz\;}
\ket{g,n-1}.
\label{eq:refrigerator_forward_cycle}
\end{equation}
The first transition absorbs energy \(\hbar\omm\) from the cold reservoir and
removes one piston quantum. The second releases energy \(\hbar\omz\) to the hot
reservoir without changing the piston occupation. Each completed cycle
therefore extracts \(\hbar\omm\) from the cold reservoir and consumes
\(\hbar\nup\) from the dressed piston mode, consistent with
\(\hbar\omz=\hbar\omm+\hbar\nup\).

The refrigerator is described by the same retained-channel TLS--piston master
equation as the heat engine, Eq.~\eqref{eq:engine_master_equation_compact}.
The change of operating mode arises from the initial conditions and the
resulting direction of the energy flow, rather than from a different
dynamical equation. As in Sec.~\ref{subsec:engine_sideband_selection}, the
finite-time Lamb shift is neglected. Under the weak-backaction product-state
approximation in Eq.~\eqref{eq:piston_level_reduction}, the joint dynamics
reduce to effective equations for the piston and the TLS. The piston therefore
obeys Eq.~\eqref{eq:effective_piston_amplifier}, with the diffusion coefficient
\(D_P(s)\) and drift coefficient \(\Gamma(s)\) defined in
Eqs.~\eqref{eq:piston_diffusion_coefficient} and
\eqref{eq:piston_drift_coefficient}, respectively. In the refrigerator
regime, positive \(\Gamma(s)\) attenuates the coherent piston amplitude and
therefore depletes the finite piston resource.
The TLS population dynamics follow from the same retained-channel master
equation~\eqref{eq:engine_master_equation_compact} by taking its excited-state matrix element and tracing over the
piston. Applying the product-state reduction in
Eq.~\eqref{eq:piston_level_reduction}, together with the rate definitions in
Eqs.~\eqref{eq:engine_hot_rates} and \eqref{eq:engine_cold_rates}, yields
\begin{align}
\dot p_e(s)
&=
-r_h^\downarrow(s)p_e(s)
+r_h^\uparrow(s)p_g(s)
+r_c^\uparrow(s)p_g(s)n_P(s)
\nonumber\\
&\qquad
-r_c^\downarrow(s)p_e(s)
\bigl[n_P(s)+1\bigr].
\label{eq:refrigerator_TLS_population_reduced}
\end{align}
Together with Eq.~\eqref{eq:piston_number_dynamics} and
\(p_g(s)=1-p_e(s)\), this provides a closed set of equations for
\(p_e(s)\) and \(n_P(s)\), which is used below to evaluate the refrigerator
dynamics and cold heat current.

The cold current follows directly from the two cold-sideband population-transfer
terms in Eq.~\eqref{eq:refrigerator_TLS_population_reduced}. The transition
\(\ket{g,n}\to\ket{e,n-1}\), which is the first step of the cooling cycle in
Eq.~\eqref{eq:refrigerator_forward_cycle}, occurs with flux
\(r_c^\uparrow(s)p_g(s)n_P(s)\) and extracts one energy quantum
\(\hbar\omm\) from the cold reservoir. The reverse transition
\(\ket{e,n-1}\to\ket{g,n}\) occurs with flux
\(r_c^\downarrow(s)p_e(s)[n_P(s)+1]\) and returns the same energy to the cold
reservoir. With \(\Jc(s)>0\) denoting heat extracted from the cold reservoir,
the net cold current is therefore
\begin{equation}
\Jc(s)
=
\hbar\omm
\left[
r_c^\uparrow(s)p_g(s)n_P(s)
-
r_c^\downarrow(s)p_e(s)
\bigl[n_P(s)+1\bigr]
\right].
\label{eq:refrigerator_cold_current}
\end{equation}
Using Eqs.~\eqref{eq:piston_diffusion_coefficient} and
\eqref{eq:piston_drift_coefficient}, this becomes
\begin{equation}
\frac{\Jc(s)}{\hbar\omm}
=
\Gamma(s)n_P(s)-D_P(s).
\label{eq:refrigerator_current_drift_relation}
\end{equation}
Comparison with Eq.~\eqref{eq:piston_number_dynamics} immediately gives
\begin{equation}
\dot n_P(s)
=
-\frac{\Jc(s)}{\hbar\omm}.
\label{eq:refrigerator_piston_depletion_identity}
\end{equation}
Thus, the same net lower sideband flux that extracts cold heat depletes the
dressed piston mode.

Refrigeration requires \(\Jc(s)>0\), which from
Eq.~\eqref{eq:refrigerator_current_drift_relation} gives
\(\Gamma(s)n_P(s)>D_P(s)\). Since \(D_P(s)\geq0\), a necessary condition is
\(\Gamma(s)>0\). The full cooling condition is therefore
\begin{align}
n_P(s)
&>
\bar n_{\min}(s),
\nonumber\\
\bar n_{\min}(s)
&=
\frac{
r_c^\downarrow(s)p_e(s)
}{
r_c^\uparrow(s)p_g(s)
-
r_c^\downarrow(s)p_e(s)
}.
\label{eq:refrigerator_threshold_occupation}
\end{align}
The quantity \(\bar n_{\min}(s)\) is the instantaneous cooling threshold.
When \(n_P(s)=\bar n_{\min}(s)\), the two lower sideband fluxes balance and
both \(\Jc(s)\) and \(\dot n_P(s)\) vanish at that instant. This threshold is
consistent with the finite-resource quantized piston refrigerator discussed
in Ref.~\cite{GelbwaserKurizkiPRE2014}.

\subsection{Anti-Zeno enhancement of finite-resource cooling}
\label{subsec:refrigerator_AZD_finite_resource}

We now compare the finite-time refrigerator dynamics with the corresponding
Markovian reference. In the finite-time calculation, the retained hot-carrier
and cold-sideband rates in Eq.~\eqref{eq:finite_time_sideband_master_equation}  are evaluated from the instantaneous rates
\(\gamma_j(\omega,s)\) in Eq.~\eqref{eq:finite_time_rate_main}, whereas the
Markovian calculation uses their long-time values
\(\gamma_j^{\rm M}(\omega)=2\pi G_j(\omega)\). Both calculations use the same
filtered reservoir spectra, system parameters, and initial state. The
difference between them therefore isolates the effect of finite-time
reservoir sampling.

\begin{figure}[!t]
    \centering

    \includegraphics[width=0.90\columnwidth]
    {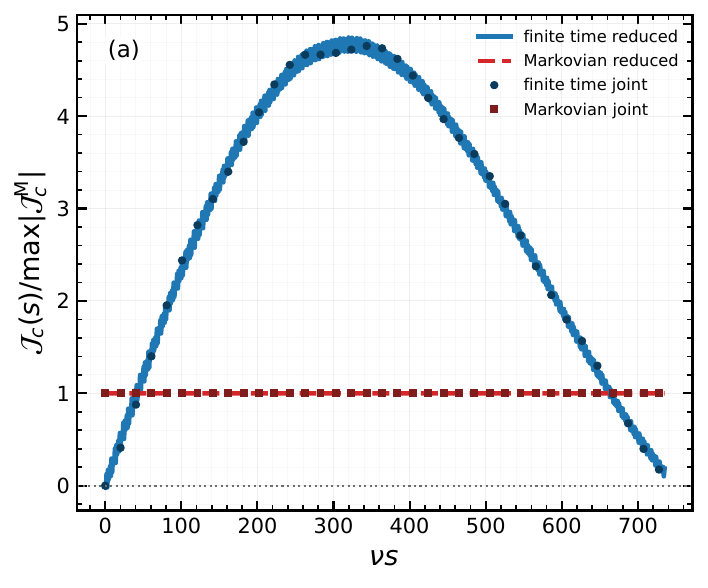}

    \vspace{0.3em}

    \includegraphics[width=0.90\columnwidth]
    {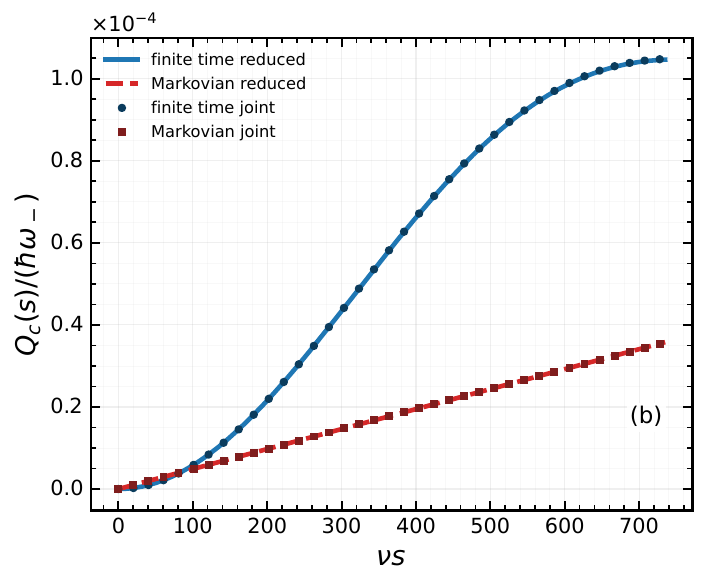}

    \caption{
    Anti-Zeno enhancement of finite-resource refrigerator operation.
    Solid blue curves are obtained by jointly solving the reduced piston-number
    and TLS-population equations,
    Eqs.~\eqref{eq:piston_number_dynamics} and
    \eqref{eq:refrigerator_TLS_population_reduced}, using the instantaneous
    finite-time rates \(\gamma_j(\omega,s)\). The red dashed curves show the
    corresponding Markovian results. Markers are obtained by independently
    propagating the retained joint TLS--piston master equation
    \eqref{eq:engine_master_equation_compact} without the product-state
    reduction, using a piston cutoff \(N_P=40\).
    Panel~(a) shows the cold current from
    Eq.~\eqref{eq:refrigerator_cold_current}, normalized by
    \(\max_s|\Jc^{\rm M}(s)|\). Panel~(b) shows the cumulative extracted cold
    heat \(Q_c(s)/(\hbar\omm)\), defined in
    Eq.~\eqref{eq:refrigerator_cumulative_cold_heat}.
    The largest current enhancement occurs near
    \(\nup s\simeq319.7\), while the trajectory is shown up to
    \(\nup s=735\), before the onset of negative retained finite-time rates.
    The filtered Lorentzian spectra are those of
    Eqs.~\eqref{eq:AZD_filtered_hot_spectrum} and
    \eqref{eq:AZD_filtered_cold_spectrum}.
    Parameters are
    \(\hbar=k_B=\nup=1\),
    \(\omz=3\nup\),
    \(\omm=2\nup\),
    \(\zeta=0.05\),
    \(\beta_h\omz=1.40\),
    \(\beta_c\omm=1.00\),
    \(n_P(0)=6\), and
    \(p_e(0)\simeq0.213982\).
    The spectra use
    \(G_{0h}=10^{-5}\),
    \(G_{0c}=8.5\times10^{-5}\),
    \(\Gamma_h=\Gamma_c=10^{-3}\nup\),
    \(\delta_h=\delta_c=5.0\times10^{-3}\nup\), and
    \(\Delta_{w,h}=\Delta_{w,c}=10^{-2}\nup\).
    }
    \label{fig:AZD_cooling_power}
\end{figure}

The reduced trajectories are obtained by solving the TLS population equation
\eqref{eq:refrigerator_TLS_population_reduced} together with the piston-number
equation~\eqref{eq:piston_number_dynamics}, with
\(p_g(s)=1-p_e(s)\). Both trajectories start from the same coherent dressed
piston state,
\begin{equation}
n_P(0)
=
|\alpha_0|^2
=
6.
\label{eq:refrigerator_initial_piston_occupation}
\end{equation}
To minimize preparation-dependent TLS transients, we choose the common initial
TLS population as the stationary solution of the retained Markovian
population dynamics at fixed \(n_P(0)\). Setting
\(\dot p_e(0)=0\) in
Eq.~\eqref{eq:refrigerator_TLS_population_reduced} with the Markovian rates
gives
\begin{equation}
p_e(0)
=
\frac{
r_{h,{\rm M}}^\uparrow
+
r_{c,{\rm M}}^\uparrow n_P(0)
}{
r_{h,{\rm M}}^\downarrow
+
r_{h,{\rm M}}^\uparrow
+
r_{c,{\rm M}}^\uparrow n_P(0)
+
r_{c,{\rm M}}^\downarrow
\bigl[n_P(0)+1\bigr]
}.
\label{eq:refrigerator_initial_TLS_population}
\end{equation}
For the parameters used below, this gives
\(p_e(0)\simeq0.213982\).

The instantaneous cooling performance is quantified by the cold current
\(\Jc(s)\) in Eq.~\eqref{eq:refrigerator_cold_current}. Its accumulated
effect is the heat extracted from the cold reservoir,
\begin{equation}
Q_c(s)
=
\int_0^s ds'\,\Jc(s').
\label{eq:refrigerator_cumulative_cold_heat}
\end{equation}
Equation~\eqref{eq:refrigerator_piston_depletion_identity} then implies
\(Q_c(s)/(\hbar\omm)=n_P(0)-n_P(s)\), so the accumulated cooling is directly
related to depletion of the finite piston resource. For the present
benchmark, the instantaneous cooling threshold in
Eq.~\eqref{eq:refrigerator_threshold_occupation} is not reached within the
propagation window over which all retained finite-time rates remain
nonnegative. We therefore quantify the anti-Zeno advantage through the
instantaneous cold current and the accumulated extracted cold heat rather
than extrapolating a threshold-crossing time.
To test the product-state reduction independently, we also propagate the
retained joint TLS--piston master equation
\eqref{eq:engine_master_equation_compact}, without imposing
\(\widetilde{\rho}(s)\simeq \rho_{\rm TLS}(s)\otimes\rho_P(s)\).
The corresponding cold current is evaluated directly from the joint
cold-sideband transition fluxes. These joint calculations provide the
numerical markers in Fig.~\ref{fig:AZD_cooling_power}, while the continuous
curves are obtained from the reduced equations above.

The anti-Zeno enhancement is seen most directly in the instantaneous cold
current in Fig.~\ref{fig:AZD_cooling_power}(a). Restricting the comparison to
the resolved regime in which the Markovian cooling current is positive and
nonvanishing, we find
\begin{equation}
\max_s
\frac{
\Jc^{\rm FT}(s)
}{
\Jc^{\rm M}(s)
}
\simeq
4.851
\label{eq:refrigerator_current_enhancement}
\end{equation}
near \(\nup s\simeq319.7\). The enhancement is transient rather than a
uniform rescaling of the refrigerator dynamics.
The integrated consequence of the enhanced cooling current is shown in
Fig.~\ref{fig:AZD_cooling_power}(b). The finite-time trajectory extracts more
heat from the cold reservoir than the Markovian reference over the same
elapsed time. At the final plotted time \(s_f\),
\begin{equation}
\frac{
Q_c^{\rm FT}(s_f)
}{
Q_c^{\rm M}(s_f)
}
\simeq
2.924,
\qquad
\nup s_f=735.
\label{eq:refrigerator_cumulative_heat_enhancement}
\end{equation}
Thus finite-time reservoir sampling yields approximately \(2.92\) times the
accumulated cold heat of the Markovian reference over this interval.

The independent joint propagation supports the product-state reduction used
for the continuous curves. The maximum sampled trace distance between the
joint TLS--piston state and the product of its reduced states is approximately
\(2.14\times10^{-4}\), and the reduced and joint cold currents differ by less
than approximately \(0.18\%\) on the global current scale. The weak-coupling
diagnostic gives
\(\gamma_{\max}\tau_B\simeq1.14\times10^{-2}\), while the
carrier-plus-first-sideband parameter in
Eq.~\eqref{eq:first_sideband_validity_condition_main} reaches
\(\max_s[2\zeta(\langle\tilde{a}^\dagger\tilde{a}\rangle_s+1)^{1/2}]\simeq0.15.\)
All retained instantaneous rates remain nonnegative over the plotted
trajectory. 

The anti-Zeno advantage is therefore a cooling-rate effect within the
retained-channel description. Finite-time reservoir sampling modifies the
transition fluxes without changing the carrier and sideband energies. Using
Eq.~\eqref{eq:refrigerator_piston_depletion_identity} together with the
dressed piston energy in Eq.~\eqref{eq:piston_total_energy}, the cold heat
extraction and piston-energy depletion satisfy
\begin{equation}
\frac{\Jc(s)}
{-\dot E_P(s)}
=
\frac{\omm}{\nup}.
\label{eq:refrigerator_channel_energy_ratio}
\end{equation}
The corresponding channel energy ratio is therefore unchanged by the
finite-time enhancement. It should be distinguished from an ergotropy-based
performance measure, since the piston energy need not remain entirely
extractable as work. The state-dependent ergotropy-based performance and the
corresponding thermal bounds are discussed in
Appendix~\ref{app:efficiency_cop}.

\section{Conclusion}
\label{sec:conclusion}

We have analyzed an autonomous thermal machine comprising a
two-level working fluid coupled to a quantized harmonic piston and two
spectrally separated thermal reservoirs. Using a polaron transformation,
we derive a finite-time master equation that resolves the reservoir-induced
dynamics into carrier and sideband transitions. Their spectral overlap
with the structured reservoirs produces an anti-Zeno regime of enhanced
energy exchange without external modulation during the operating interval.
For an initially coherent piston, this enhancement increases the ergotropy
generated relative to the corresponding Markovian evolution over the same
interval, thereby increasing the average ergotropy charging power.
In the refrigeration regime, the same retained channels
yield a larger cooling current and greater accumulated cold-heat extraction
during consumption of the finite piston resource.

The thermodynamic interpretation follows directly from the retained channel
structure. Finite-time reservoir sampling changes the transition rates and
event fluxes but not the carrier and sideband energies. The completed-cycle
engine channel ratio is therefore fixed by \(\nup/\omz\), with the operational
energy efficiency approaching this value when the TLS remains stationary on
the piston time scale. For refrigeration, the same lower-sideband flux governs
cold-heat extraction and dressed-piston energy depletion, giving the channel
ratio \(\omm/\nup\). These ratios satisfy the usual Carnot bounds in the
Markovian thermal limit and remain consistent with them over the finite-time
windows considered here. They should nevertheless be distinguished from
ergotropy-based performance measures. The state comparison in
Appendix~\ref{app:piston_state_ergotropy} demonstrates why this distinction
matters. For the preparations and parameters considered, finite-time
dynamics enhances coherent-piston charging but accelerates ergotropy loss
for squeezed and Fock inputs, while a thermal input remains passive.
Enhanced energy exchange therefore does not imply a universal
useful-work benefit.

An experimental realization would test whether the finite-time response of an engineered thermal environment
can increase the extractable work accumulated by an autonomous quantum piston. Measuring the piston energy together with its
ergotropy would distinguish useful charging from passive energy accumulation, while varying the initial preparation would test the predicted state
dependence. Comparing measurements over different operating intervals with the Markovian prediction obtained from independently characterized reservoirs
would assess the role of finite-time reservoir response. Such a demonstration would provide an experimental basis for choosing reservoir spectra, piston
preparations, and operating intervals to improve autonomous work storage.

One possible implementation of our scheme is based on superconducting circuit QED. A superconducting qubit can serve as the TLS, while a high-\(Q\)
microwave resonator provides the quantized piston and frequency-selective circuit elements engineer the hot and cold
reservoirs~\cite{GelbwaserKurizkiPRE2014,Ronzani2018HeatValve,Aamir2025}. Quantum-load characterization and ergotropy evaluation have
already been demonstrated in trapped-ion engines~\cite{vonLindenfels2019,VanHorne2020}. Relevant circuit capabilities
include qubit--oscillator coupling, photonic heat transport, and autonomous refrigeration~\cite{Bera2021Longitudinal,Ronzani2018HeatValve,Aamir2025},
together with a recent work reporting an autonomous superconducting heat engine~\cite{Uusnakki2026Autonomous}. Superconducting-qubit experiments
have also demonstrated suppression and enhancement of decay in structured environments~\cite{Harrington2017Zeno,Thorbeck2024ZenoAntiZeno}.
Those experiments employ measurement or readout, whereas the present proposal concerns evolution without such interventions during the operating
interval. These developments support the relevant observables and constituent capabilities.

The present results are obtained within weak TLS--reservoir coupling, resolved sidebands, and the carrier-plus-first-sideband truncation. Extending the
analysis beyond these regimes could reveal how stronger coupling, higher sidebands, and more pronounced reservoir memory modify the autonomous
machine. Exploring a broader range of piston preparations and reservoir spectra would clarify the scope of the charging benefit
beyond the cases examined here.

\begin{acknowledgments}
M. Tahir Naseem acknowledges support from the National Center for Quantum Computing (NCQC), Pakistan.
\end{acknowledgments}

\section*{Data Availability}
The numerical codes used to generate the results presented in this work are available in Ref.~\cite{DataRepository}. 

\appendix

\section{Derivation of the finite-time sideband master equation}
\label{app:finite_time_master_equation}

This appendix derives the finite-time sideband master equation
\eqref{eq:finite_time_sideband_master_equation}. We use the time convention
introduced in Sec.~\ref{subsec:finite_time_sideband_master_equation}. Here,
\(s\) denotes the elapsed evolution time, \(s'\) an earlier time, and
\(u=s-s'\) the reservoir-memory delay. The quantity \(\tau_c\) denotes the
selected endpoint at which the evolution is evaluated, with
\(0\leq s\leq\tau_c\). No switching operation is implied at \(s=\tau_c\).
The derivation proceeds by resolving the dressed interaction into carrier and
first-sideband components, constructing the finite-time TCL2 generator,
evaluating the thermal reservoir response, and finally applying the secular approximation.

Applying the first-order displacement expansion in
Eq.~\eqref{eq:first_sideband_expansion_main} to the dressed interaction in
Eq.~\eqref{eq:polaron_transformed_HSB_main} gives the carrier and
first-sideband contributions. Transforming the TLS-lowering part to the
interaction picture generated by the system Hamiltonian in
Eq.~\eqref{eq:polaron_system_hamiltonian_main} yields
\begin{align}
\widetilde{S}_{I,\downarrow}(s)
\simeq{}&
\tilde{\sigma}_-e^{-i\omz s}
+
2\zeta
\left(
\tilde{\sigma}_-\tilde{a}\,
e^{-i\omp s}
-
\tilde{\sigma}_-\tilde{a}^\dagger\,
e^{-i\omm s}
\right).
\label{eq:app_lowering_system_operator_IP}
\end{align}
The first-order expansion remains valid provided the occupation-dependent
weak-coupling condition in Eq.~\eqref{eq:first_sideband_validity_condition_main} is satisfied. The dressed piston
operators \(\tilde{a}\) and \(\tilde{a}^{\dagger}\) are defined in
Eq.~\eqref{eq:polaron_ladder_operators_main}, while the dressed TLS transition
operators \(\tilde{\sigma}_{\pm}\) are related to their bare counterparts by
Eq.~\eqref{eq:polaron_sigma_pm_main}. Taking the Hermitian conjugate of
Eq.~\eqref{eq:app_lowering_system_operator_IP} gives the corresponding
upward-transition contribution. The complete system coupling operator is
therefore
\begin{equation}
\widetilde{S}_I(s)
=
\widetilde{S}_{I,\downarrow}(s)
+
\widetilde{S}_{I,\downarrow}^{\dagger}(s).
\label{eq:app_system_coupling_operator_decomposition}
\end{equation}
For the frequency-domain calculation, we extend the positive-frequency set in
Eq.~\eqref{eq:sideband_frequency_set_main} to
\begin{equation}
\Omega_{\rm sb}^{\pm}
=
\{
\pm\omz,\pm\omm,\pm\omp
\},
\quad
\widetilde{A}(-\omega)
=
\widetilde{A}^\dagger(\omega),
\quad
\omega\in\Omega_{\rm sb}.
\label{eq:app_signed_frequency_convention}
\end{equation}
This signed-frequency convention allows upward and downward transition
components to be treated within a single frequency sum. The
positive-frequency transition operators \(\widetilde{A}(\omega)\) are defined
in Eq.~\eqref{eq:sideband_transition_operators_main}. After secularization,
the signed-frequency contributions will be reorganized into separate upward
and downward terms.
Using the decomposition in
Eq.~\eqref{eq:app_system_coupling_operator_decomposition}, the dressed
TLS--reservoir interaction in
Eq.~\eqref{eq:polaron_transformed_HSB_main} takes the following form in the
interaction picture:
\begin{equation}
\widetilde{H}_I(s)
=
\sum_{j=h,c}
\widetilde{S}_I(s)\otimes B_j(s),
\label{eq:app_interaction_hamiltonian_IP}
\end{equation}
where
\begin{equation}
\widetilde{S}_I(s)
=
\sum_{\omega\in\Omega_{\rm sb}^{\pm}}
\widetilde{A}(\omega)e^{-i\omega s}.
\label{eq:app_system_operator_IP}
\end{equation}
The interaction-picture reservoir operator is
\(B_j(s)=e^{iH_{B_j}s/\hbar}B_je^{-iH_{B_j}s/\hbar}\).

We now derive the reduced dynamics to second order in the TLS--reservoir
interaction. The TCL2 treatment assumes weak TLS--reservoir coupling, an
initially factorized state
\(\widetilde{\rho}(0)\otimes\rho_B\), and two independent stationary
reservoirs with
\(\rho_B=\rho_{B_h}\otimes\rho_{B_c}\). We further assume vanishing reservoir
means,
\(\Tr_{B_j}[B_j\rho_{B_j}]=0\), so that the first-order contribution
vanishes. Under these assumptions, the second-order time-convolutionless
expansion gives~\cite{BreuerPetruccione2002}
\begin{equation}
\begin{aligned}
\frac{d\widetilde{\rho}_I(s)}{ds}
=
-\frac{1}{\hbar^2}
\int_0^s ds'\,
\Tr_B
\Big[
\widetilde{H}_I(s),
\big[
\widetilde{H}_I(s'),
\widetilde{\rho}_I(s)\otimes\rho_B
\big]
\Big].
\end{aligned}
\label{eq:app_TCL2_start}
\end{equation}
The occurrence of \(\widetilde{\rho}_I(s)\), rather than
\(\widetilde{\rho}_I(s')\), makes Eq.~\eqref{eq:app_TCL2_start} local in the
elapsed time.

To retain the finite-memory effects, we introduce the stationary correlation
function of reservoir \(j\),
\begin{equation}
C_j(u)
=
\frac{1}{\hbar^2}
\Tr_{B_j}
\left[
B_j(u)B_j(0)\rho_{B_j}
\right],
\label{eq:bath_correlation_main}
\end{equation}
which measures reservoir correlations between two times separated by the
memory delay \(u\). Its finite-time Fourier transform defines the response
coefficient
\begin{equation}
\Gamma_j(\omega,s)
=
\int_0^s du\,
e^{i\omega u}C_j(u).
\label{eq:finite_time_response_main}
\end{equation}
Retaining the finite upper limit \(s\), rather than extending it to infinity,
keeps the response explicitly time dependent. Its real and imaginary parts
determine the finite-time dissipative rate and Lamb-shift coefficient,
respectively. In addition, independence of the two reservoirs eliminates their cross-correlations,
\begin{equation}
\Tr_B
\left[
B_j(s)B_\ell(s-u)\rho_B
\right]
=
\delta_{j\ell}\hbar^2 C_j(u).
\label{eq:app_independent_bath_correlation}
\end{equation}

Substituting Eqs.~\eqref{eq:app_interaction_hamiltonian_IP} and
\eqref{eq:app_system_operator_IP} into
Eq.~\eqref{eq:app_TCL2_start}, and using
Eq.~\eqref{eq:app_independent_bath_correlation}, gives the frequency-resolved
pre-secular generator. Here the sums over \(\omega\) and \(\omega'\) run over
the signed-frequency set \(\Omega_{\rm sb}^{\pm}\):
\begin{align}
\frac{d\widetilde{\rho}_I(s)}{ds}
={}&
\sum_{j=h,c}
\sum_{\omega,\omega'}
e^{i(\omega'-\omega)s}
\Gamma_j(\omega,s)
\left[
\widetilde{A}(\omega)\widetilde{\rho}_I(s),
\widetilde{A}^\dagger(\omega')
\right]
\nonumber\\
&+
\mathrm{H.c.}.
\label{eq:app_presecular_generator}
\end{align}
The phase factor \(e^{i(\omega'-\omega)s}\) explicitly identifies the cross
terms between distinct signed transition frequencies. These terms are
retained at this stage and removed only when the secular approximation is applied.

We next evaluate the response coefficient \(\Gamma_j(\omega,s)\) in Eq.~\eqref{eq:app_presecular_generator} for the
bosonic thermal reservoirs specified in
Sec.~\ref{subsec:microscopic_setup}. From
Eqs.~\eqref{eq:bath_hamiltonian} and~\eqref{eq:TLS_bath_coupling}, the
interaction-picture reservoir operator is
\begin{equation}
B_j(s)
=
\sum_k
\hbar\lambda_{jk}
\left(
b_{jk}e^{-i\omega_{jk}s}
+
b_{jk}^\dagger e^{i\omega_{jk}s}
\right).
\label{eq:app_bath_operator_time}
\end{equation}
Because reservoir \(j\) is in a thermal state at inverse temperature
\(\beta_j\), its nonzero second moments are
\begin{equation}
\langle b_{jk}^\dagger b_{jk}\rangle
=
n_j(\omega_{jk}),
\qquad
\langle b_{jk}b_{jk}^\dagger\rangle
=
n_j(\omega_{jk})+1,
\label{eq:app_thermal_averages}
\end{equation}
where
\begin{equation}
n_j(\omega)
=
\frac{1}{e^{\beta_j\hbar\omega}-1}
\label{eq:app_bose_factor}
\end{equation}
is the Bose occupation factor. Substituting
Eqs.~\eqref{eq:app_bath_operator_time} and
\eqref{eq:app_thermal_averages} into
Eq.~\eqref{eq:bath_correlation_main} gives
\begin{equation}
\begin{aligned}
C_j(u)
=
\sum_k
|\lambda_{jk}|^2
\Big[
&
\big(n_j(\omega_{jk})+1\big)e^{-i\omega_{jk}u}
+
n_j(\omega_{jk})e^{i\omega_{jk}u}
\Big].
\end{aligned}
\label{eq:app_correlation_discrete}
\end{equation}
We introduce the positive-frequency spectral density
\begin{equation}
J_j(\omega)
=
\sum_k
|\lambda_{jk}|^2
\delta(\omega-\omega_{jk}),
\qquad
\omega>0.
\label{eq:app_spectral_density}
\end{equation}
The thermal occupation factors can then be incorporated into a two-sided
response spectrum defined, for \(\omega>0\), by
\begin{equation}
G_j(\omega)
=
J_j(\omega)\big[n_j(\omega)+1\big],
\qquad
G_j(-\omega)
=
J_j(\omega)n_j(\omega).
\label{eq:app_two_sided_spectrum}
\end{equation}
Thus, \(J_j(\omega)\) specifies the positive-frequency reservoir spectral
density, whereas \(G_j(\xi)\) incorporates the thermal factors into the
positive- and negative-frequency branches. Using Eq.~\eqref{eq:app_two_sided_spectrum}, the discrete correlation function
in Eq.~\eqref{eq:app_correlation_discrete} then takes the compact form
\begin{equation}
C_j(u)
=
\int_{-\infty}^{\infty}
d\xi\,
G_j(\xi)e^{-i\xi u}.
\label{eq:app_correlation_twosided}
\end{equation}
The ratio \(n_j(\omega)/[n_j(\omega)+1]\) gives the thermal
detailed-balance relation
\begin{equation}
G_j(-\omega)
=
e^{-\beta_j\hbar\omega}G_j(\omega),
\qquad
\omega>0.
\label{eq:app_detailed_balance}
\end{equation}

Substituting the two-sided representation in
Eq.~\eqref{eq:app_correlation_twosided} into the finite-time response
coefficient in Eq.~\eqref{eq:finite_time_response_main}, and carrying out the
\(u\)-integration, gives
\begin{align}
\Gamma_j(\omega,s)
&=
\int_{-\infty}^{\infty}
d\xi\,
G_j(\xi)
\frac{
e^{i(\omega-\xi)s}-1
}{
i(\omega-\xi)
}.
\label{eq:app_Gamma_spectral}
\end{align}
Writing
\(
\Gamma_j(\omega,s)
=
\gamma_j(\omega,s)/2+i\delta_j^{\rm LS}(\omega,s)
\),
the real part gives
\begin{equation}
\gamma_j(\omega,s)
=
2
\int_{-\infty}^{\infty}
d\xi\,
G_j(\xi)
\frac{
\sin[(\omega-\xi)s]
}{
\omega-\xi
},
\label{eq:app_finite_time_rate}
\end{equation}
which reproduces the finite-time rate in
Eq.~\eqref{eq:finite_time_rate_main}. The imaginary part of Eq.~\eqref{eq:app_Gamma_spectral} gives the
corresponding finite-time Lamb-shift coefficient,
\begin{equation}
\delta_j^{\rm LS}(\omega,s)
=
\int_{-\infty}^{\infty}
d\xi\,
G_j(\xi)
\frac{
1-\cos[(\omega-\xi)s]
}{
\omega-\xi
}.
\label{eq:app_finite_time_lamb}
\end{equation}

We now apply the secular approximation to the pre-secular generator
in Eq.~\eqref{eq:app_presecular_generator}. Terms with
\(\omega\neq\omega'\) are neglected when the distinct signed transition
frequencies remain spectrally resolved, leaving only
\(\omega=\omega'\). Using the signed-frequency convention in
Eq.~\eqref{eq:app_signed_frequency_convention} together with the dissipative
and Lamb-shift coefficients in Eqs.~\eqref{eq:app_finite_time_rate} and
\eqref{eq:app_finite_time_lamb}, the remaining terms can be reorganized over
the positive-frequency set \(\Omega_{\rm sb}\). Transforming back to the
Schrödinger picture, Eq.~\eqref{eq:app_presecular_generator} gives
\begin{align}
\frac{d\widetilde{\rho}(s)}{ds}
={}&
-\frac{i}{\hbar}
\left[
\widetilde{H}_{\rm sys}
+
\widetilde{H}_{\rm LS}(s),
\widetilde{\rho}(s)
\right]
\nonumber\\
&+
\sum_{j=h,c}
\sum_{\omega\in\Omega_{\rm sb}}
\Big[
\gamma_j(\omega,s)
\D[\widetilde{A}(\omega)]\widetilde{\rho}(s)
\nonumber\\
&\hspace{1.0cm}
+
\gamma_j(-\omega,s)
\D[\widetilde{A}^\dagger(\omega)]\widetilde{\rho}(s)
\Big].
\label{eq:app_compact_sideband_master_equation}
\end{align}
Here, the finite-time Lamb-shift Hamiltonian is
\begin{equation}
\begin{aligned}
\widetilde{H}_{\rm LS}(s)
=
\hbar
\sum_{j=h,c}
\sum_{\omega\in\Omega_{\rm sb}}
\Big[
&
\delta_j^{\rm LS}(\omega,s)
\widetilde{A}^\dagger(\omega)\widetilde{A}(\omega)
\\
&+
\delta_j^{\rm LS}(-\omega,s)
\widetilde{A}(\omega)\widetilde{A}^\dagger(\omega)
\Big].
\end{aligned}
\label{eq:app_lamb_shift}
\end{equation}
Equation~\eqref{eq:app_compact_sideband_master_equation} reproduces the
finite-time sideband master equation in Eq.~\eqref{eq:finite_time_sideband_master_equation}.

Finally, the long-time limit provides a consistency check. For times long
compared with the reservoir correlation time, the kernel in
Eq.~\eqref{eq:app_finite_time_rate} approaches, in the distributional sense,
\begin{equation}
\frac{
\sin[(\omega-\xi)s]
}{
\omega-\xi
}
\longrightarrow
\pi\delta(\omega-\xi),
\qquad
s\rightarrow\infty.
\label{eq:app_markov_delta_limit}
\end{equation}
It follows that
\(\gamma_j(\omega,s)\rightarrow2\pi G_j(\omega)\), and the finite-time
dissipative rates reduce to the standard weak-coupling Markovian golden-rule
rates.

\section{State-resolved piston ergotropy}
\label{app:piston_state_ergotropy}

The main text in Sec.~\ref{sec:quantum_heat_engine} uses an initially coherent piston as the engine benchmark, for
which the ergotropy is carried entirely by the coherent displacement. Here we
compare coherent, squeezed-vacuum, Fock, and thermal piston preparations under
the same reduced piston dynamics derived in
Sec.~\ref{subsec:piston_amplification_ergotropy}. The purpose is to isolate how
the initial piston state affects the conversion of channel gain into extractable
work. Accordingly, the amplifier coefficients are held fixed across the four
preparations. 

{\it{Common finite-time amplifier and Gaussian ergotropy}}: We start from the phase-insensitive piston amplifier in
Eq.~\eqref{eq:effective_piston_amplifier} and use the gain coefficient
\(\Lambda(s)=-\Gamma(s)\) introduced in
Sec.~\ref{subsec:AZD_piston_ergotropy}. The two dissipative coefficients in
Eq.~\eqref{eq:effective_piston_amplifier} are
therefore \(D_P(s)\) and \(D_P(s)-\Lambda(s)\). We restrict the comparison to
the parameter window in which both are nonnegative.

Using the coherent amplitude \(\alpha(s)\) defined in
Eq.~\eqref{eq:piston_amplitude_equation}, we introduce the centered
second moments
\begin{align}
n_c(s)
&=
\langle\tilde a^\dagger\tilde a\rangle_s
-
|\alpha(s)|^2,
\nonumber\\
m_c(s)
&=
\langle\tilde a^2\rangle_s
-
\alpha(s)^2.
\end{align}
Together with Eq.~\eqref{eq:effective_piston_amplifier} and Eq.~\eqref{eq:piston_amplitude_equation}, the reduced amplifier gives
\begin{align}
\dot n_c(s)
&=
D_P(s)+\Lambda(s)n_c(s),
\nonumber\\
\dot m_c(s)
&=
\Lambda(s)m_c(s).
\label{eq:app_state_centered_moment_dynamics}
\end{align}
It is convenient to introduce the cumulative gain
\begin{equation}
\mathcal{G}(s)
=
\exp\!\left[
\int_0^s ds'\,\Lambda(s')
\right],
\label{eq:app_state_cumulative_gain}
\end{equation}
and the diffusion-added occupation
\begin{equation}
\mathcal{N}(s)
=
\mathcal{G}(s)
\int_0^s ds'\,
\frac{D_P(s')}{\mathcal{G}(s')}.
\label{eq:app_state_added_noise}
\end{equation}
The solutions of Eq.~\eqref{eq:app_state_centered_moment_dynamics} are therefore
\begin{align}
\alpha(s)
&=
\sqrt{\mathcal{G}(s)}\,\alpha_0,
\nonumber\\
n_c(s)
&=
\mathcal{G}(s)n_c(0)+\mathcal{N}(s),
\nonumber\\
m_c(s)
&=
\mathcal{G}(s)m_c(0).
\label{eq:app_state_moment_maps}
\end{align}
Correspondingly, the total mean occupation satisfies
\(n_P(s)=\mathcal{G}(s)n_P(0)+\mathcal{N}(s)\). Thus the phase-insensitive
diffusion \(\mathcal{N}(s)\) enters additively only through the isotropic occupation. The coherent
displacement \(\alpha(s)\) and anomalous moment \(m_c(s)\) are instead governed by the cumulative gain.

The coherent, squeezed-vacuum, and thermal inputs remain Gaussian under the
linear phase-insensitive channel given in Eq.~\eqref{eq:effective_piston_amplifier}. For a single-mode Gaussian state, the passive
state associated with its spectrum is thermal, with occupation fixed by the
symplectic eigenvalue~\cite{BrownFriisHuber2016}
\begin{equation}
\vartheta(s)
=
\sqrt{
\left[n_c(s)+\frac12\right]^2
-
|m_c(s)|^2
}.
\label{eq:app_state_symplectic}
\end{equation}
With the piston Hamiltonian convention of
Eq.~\eqref{eq:bare_TLS_piston_hamiltonian}, the corresponding passive
occupation is \(\vartheta(s)-1/2\). Using the piston-energy and ergotropy
definitions in Eqs.~\eqref{eq:piston_total_energy} and
\eqref{eq:piston_ergotropy_definition}, together with
\(\langle\tilde a^\dagger\tilde a\rangle_s
=|\alpha(s)|^2+n_c(s)\) and the symplectic eigenvalue in
Eq.~\eqref{eq:app_state_symplectic}, the ergotropy therefore becomes
\begin{equation}
\WP(s)
=
\hbar\nup
\left[
|\alpha(s)|^2
+n_c(s)
-\vartheta(s)
+\frac12
\right].
\label{eq:app_state_gaussian_ergotropy}
\end{equation}
The zero-point energy is omitted from the piston Hamiltonian and would cancel
from the ergotropy in any case. Equation~\eqref{eq:app_state_gaussian_ergotropy}
applies to single-mode Gaussian piston states. For a general, including non-Gaussian, state
the ergotropy is instead evaluated from the passive-state definition in Eq.~\eqref{eq:piston_ergotropy_definition}.

{\it{Coherent, squeezed, and thermal inputs}}: For an initially coherent piston,
\(n_c(0)=m_c(0)=0\). Equations
\eqref{eq:app_state_moment_maps} and \eqref{eq:app_state_symplectic} give
\(n_c(s)=\mathcal{N}(s)\), \(m_c(s)=0\), and
\(\vartheta(s)=\mathcal{N}(s)+1/2\). Hence, using Eqs.~\eqref{eq:app_state_gaussian_ergotropy} and
\eqref{eq:piston_amplitude_solution}, the coherent-state ergotropy becomes
\begin{equation}
\WP^{\rm coh}(s)
=
\hbar\nup|\alpha_0|^2\mathcal{G}(s),
\label{eq:app_state_coherent_ergotropy}
\end{equation}
which is consistent with Eq.~\eqref{eq:coherent_state_ergotropy}. 
For a squeezed-vacuum input \(S(r,\phi)\ket{0}\), the displacement vanishes and
\[
n_c(0)=\sinh^2r,
\qquad
|m_c(0)|=\sinh r\cosh r.
\]
The phase of \(m_c(0)\) does not affect the ergotropy. Equation
\eqref{eq:app_state_moment_maps} gives
\[
n_c(s)
=
\mathcal{G}(s)\sinh^2r+\mathcal{N}(s),
\quad
|m_c(s)|
=
\mathcal{G}(s)\sinh r\cosh r,
\]
and therefore
\begin{equation}
\vartheta_{\rm sq}(s)
=
\sqrt{
\left[n_c(s)+\frac{1}{2}\right]^2
-
|m_c(s)|^2
}.
\label{eq:app_state_squeezed_symplectic}
\end{equation}
Using Eq.~\eqref{eq:app_state_gaussian_ergotropy} with \(\alpha(s)=0\), the
squeezed-state ergotropy is
\begin{equation}
\WP^{\rm sq}(s)
=
\hbar\nup
\left[
\mathcal{G}(s)\sinh^2r
+
\mathcal{N}(s)
-
\vartheta_{\rm sq}(s)
+
\frac{1}{2}
\right].
\label{eq:app_state_squeezed_ergotropy}
\end{equation}
The squeezed-state ergotropy is therefore determined by the competition between
the amplified quadrature anisotropy and the isotropic occupation added by the
phase-insensitive channel.

For a thermal input with mean occupation \(\bar n_{\rm th}\),
\(\alpha(0)=m_c(0)=0\) and \(n_c(0)=\bar n_{\rm th}\). The channel preserves
\(m_c(s)=0\) and gives
\[
n_c(s)
=
\mathcal{G}(s)\bar n_{\rm th}
+
\mathcal{N}(s).
\]
Thus \(\vartheta(s)=n_c(s)+1/2\) and
using Eq.~\eqref{eq:app_state_gaussian_ergotropy}, the ergotropy for a thermal input state becomes \(\WP^{\rm th}(s)=0\) for every coupling time. The thermal piston can therefore
gain mean energy while remaining passive.

\begin{figure}[t]
    \centering
    \includegraphics[width=\linewidth]{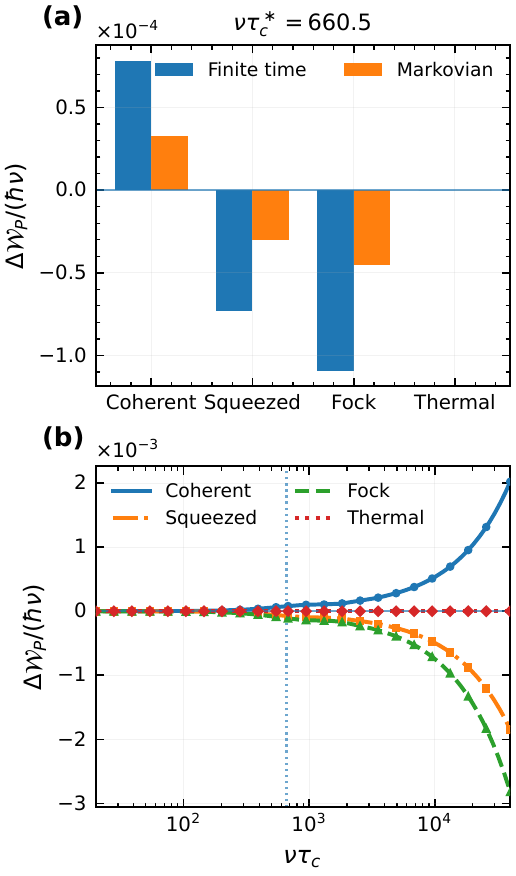}
    \caption{
    State-resolved piston ergotropy change
    \(\Delta\WP(\tau_c)=\WP(\tau_c)-\WP(0)\) for coherent
    \((|\alpha_0|^2=1)\), squeezed-vacuum \((\sinh^2r=1)\), Fock
    \((m=1)\), and thermal \((\bar n_{\rm th}=1)\) inputs.
    The microscopic reservoir parameters are identical to those of
    Fig.~\ref{fig:AZD_piston_ergotropy}.
    (a) Finite-time and Markovian ergotropy changes near the anti-Zeno gain
    maximum, \(\nup\tau_c^\ast\simeq6.6\times10^2\).
    (b) Finite-time ergotropy change versus \(\nup\tau_c\) at the physical
    reservoir scale. Solid curves show the analytical results, while markers
    denote independent numerical verification obtained from direct
    density-matrix propagation of the gain--loss master equation Eq.~\eqref{eq:effective_piston_amplifier}.
    The vertical dotted line marks \(\tau_c^\ast\).
    }
    \label{fig:state_resolved_ergotropy}
\end{figure}

{\it{Fock input}}: For a Fock input \(\ket{m}\), the Gaussian expression
Eq.~\eqref{eq:app_state_gaussian_ergotropy} does not apply. Since the input is
diagonal in the number basis and the amplifier is phase-insensitive, the piston
state remains diagonal,
\[
\rho_P(s)
=
\sum_{n=0}^{\infty}
p_n(s)\ket{n}\bra{n}.
\]
The populations obey
\begin{align}
\dot p_n(s)
={}&
D_P(s)
\left[
n p_{n-1}(s)
-
(n+1)p_n(s)
\right]
\nonumber\\
&+
\left[D_P(s)-\Lambda(s)\right]
\left[
(n+1)p_{n+1}(s)
-
n p_n(s)
\right],
\label{eq:app_state_fock_birthdeath}
\end{align}
with \(p_{-1}(s) = 0\). The first line describes piston-quantum addition and
the second piston-quantum removal. If \(p_n^\downarrow(s)\) denotes the
probabilities rearranged in decreasing order over increasing oscillator energy,
the ergotropy is
\begin{equation}
\WP^{\rm Fock}(s)
=
\hbar\nup
\left[
\sum_{n=0}^{\infty}
n p_n(s)
-
\sum_{n=0}^{\infty}
n p_n^\downarrow(s)
\right].
\label{eq:app_state_fock_ergotropy}
\end{equation}
At \(s=0\), the passive rearrangement of \(\ket{m}\bra{m}\) is the ground
state, so \(\WP^{\rm Fock}(0)=m\hbar\nup\). Under the present benchmark, the
birth-death dynamics in Eq.~\eqref{eq:app_state_fock_birthdeath} broadens the number distribution and progressively reduces
its nonpassive population ordering. For the state comparison below, we use the ergotropy change
\(\Delta\WP(s)=\WP(s)-\WP(0)\). This quantity resolves the small
state-dependent work changes while keeping the microscopic reservoir coupling
identical to the main engine benchmark.

{\it{State comparison}}: Figure~\ref{fig:state_resolved_ergotropy} compares the four initial piston
states under the same physical reservoir parameters. The finite-time curves use
the instantaneous coefficients entering Eqs.~\eqref{eq:app_state_cumulative_gain}
and~\eqref{eq:app_state_added_noise}, whereas the Markovian comparison in
panel~(a) uses their long-time golden-rule limits. 
At the anti-Zeno point in Fig.~\ref{fig:state_resolved_ergotropy}(a), the
coherent piston has a positive ergotropy change, and the finite-time result
exceeds the Markovian result by a factor \(\simeq2.39\), consistent with the
gain enhancement in Sec.~\ref{subsec:AZD_piston_ergotropy}. The squeezed and
Fock inputs instead have negative ergotropy changes at the same physical time, and
the finite-time channel increases the magnitude of this loss relative to the
Markovian channel. The thermal input remains passive and therefore has
\(\Delta\WP=0\). Thus enhancement of the finite-time amplifier does not by
itself imply enhanced ergotropy for every piston preparation.

Figure~\ref{fig:state_resolved_ergotropy}(b) shows the finite-time evolution
over the same long-time window used for the engine benchmark in Fig.~\ref{fig:AZD_piston_ergotropy}. For the present
parameters, the coherent ergotropy increases monotonically, whereas the
squeezed and Fock ergotropies decrease monotonically from their initial values.
For the squeezed input, the isotropic added occupation increases the passive
Gaussian contribution relative to the quadrature anisotropy, while the Fock-state
decrease reflects broadening of the nonpassive number distribution. The thermal
state remains passive throughout. The comparison therefore clarifies the role of the coherent-state assumption in
the main engine analysis presented in Sec.~\ref{sec:quantum_heat_engine}. A coherent piston converts the finite-time gain
directly into displacement ergotropy, whereas for more general initial states
the same phase-insensitive dynamics can leave the ergotropy unchanged, increase
it, or reduce it.

\section{Channel energy ratios and ergotropy-based performance}
\label{app:efficiency_cop}

This appendix clarifies the energy-flow and useful-work measures associated
with the retained hot-carrier and cold-lower-sideband dynamics. The reduced
piston energy \(E_P(s)\) and ergotropy \(\WP(s)\) are defined in
Eqs.~\eqref{eq:piston_total_energy} and
\eqref{eq:piston_ergotropy_definition}, respectively. The general
thermodynamic distinction between piston energy and extractable work, and the
corresponding state-dependent performance bounds for autonomous
quantized-piston machines, were developed in
Ref.~\cite{GelbwaserKurizkiPRE2014}. Here we specialize this distinction to
the finite-time retained-channel dynamics considered in the main text. 

For the retained dissipative dynamics, we take \(\dot Q_j(s)>0\) to denote heat
flowing from reservoir \(j=h,c\) into the TLS--piston system. Since
\(\dot E_{\rm TLS}(s)=\hbar\omz\dot p_e(s)\), energy conservation gives
\begin{equation}
\dot E_{\rm TLS}(s)
+
\dot E_P(s)
=
\dot Q_h(s)
+
\dot Q_c(s).
\label{eq:app_first_law_sign_convention}
\end{equation}
When the TLS remains approximately stationary on the slower piston time scale,
\(\dot E_{\rm TLS}(s)\simeq0\), Eq.~\eqref{eq:app_first_law_sign_convention}
reduces to the familiar autonomous-piston energy balance discussed in
Ref.~\cite{GelbwaserKurizkiPRE2014}. For a quantized piston, the total energy
change need not coincide with the change in useful work content. From
Eq.~\eqref{eq:piston_ergotropy_definition},
\begin{equation}
\dot E_P(s)
=
\dot{\WP}(s)
+
\dot E_{\rm pas}(s),
\label{eq:app_piston_energy_ergotropy_split}
\end{equation}
where \(E_{\rm pas}(s)\) denotes the passive part of the piston energy. This
decomposition distinguishes the channel-energy ratios below from
state-dependent ergotropy-based performance.

To describe both operating modes with a common convention, we define the net
hot-carrier excitation flux and cold-lower-sideband relaxation flux as
\begin{align}
\Phi_h(s)
&=
r_h^\uparrow(s)p_g(s)
-
r_h^\downarrow(s)p_e(s),
\nonumber\\
\Phi_c(s)
&=
r_c^\downarrow(s)p_e(s)
\bigl[n_P(s)+1\bigr]
-
r_c^\uparrow(s)p_g(s)n_P(s).
\label{eq:app_channel_event_fluxes}
\end{align}
Positive \(\Phi_h\) and \(\Phi_c\) correspond to the forward engine cycle in
Eq.~\eqref{eq:engine_forward_cycle}. The TLS population equation~\eqref{eq:refrigerator_TLS_population_reduced} then takes
the compact form
\begin{equation}
\dot p_e(s)
=
\Phi_h(s)
-
\Phi_c(s).
\label{eq:app_population_flux_balance}
\end{equation}
Using the carrier and lower-sideband transition energies in
Eq.~\eqref{eq:sideband_frequency_set_main}, the associated energy currents are
\begin{align}
\dot Q_h(s)
&=
\hbar\omz\,\Phi_h(s),
\nonumber\\
\dot Q_c(s)
&=
-\hbar\omm\,\Phi_c(s),
\nonumber\\
\dot E_P(s)
&=
\hbar\nup\,\Phi_c(s).
\label{eq:app_channel_energy_currents}
\end{align}
Equations~\eqref{eq:app_population_flux_balance} and
\eqref{eq:app_channel_energy_currents} reproduce the exact balance in
Eq.~\eqref{eq:app_first_law_sign_convention}, since
\(\omz=\omm+\nup\).

For engine operation, one completed forward cycle absorbs
\(\hbar\omz\) from the hot reservoir and transfers
\(\hbar\nup\) to the piston. The corresponding channel-energy ratio is
therefore
\begin{equation}
\eta_{\rm ch}
=
\frac{\nup}{\omz}
=
1-\frac{\omm}{\omz}.
\label{eq:app_engine_channel_efficiency}
\end{equation}
This frequency-ratio description is consistent with the thermodynamics of
autonomous quantized-piston machines developed in
Ref.~\cite{GelbwaserKurizkiPRE2014}. During finite-time evolution, however,
the TLS can temporarily store energy. The instantaneous piston-energy ratio is
therefore
\begin{equation}
\eta_E(s)
=
\frac{\dot E_P(s)}
{\dot Q_h(s)}
=
\eta_{\rm ch}
\frac{\Phi_c(s)}
{\Phi_h(s)},
\label{eq:app_engine_operational_energy_efficiency}
\end{equation}
whenever \(\dot Q_h(s)>0\) and \(\dot E_P(s)>0\).
Equation~\eqref{eq:app_population_flux_balance} shows that
\(\Phi_h(s)\simeq\Phi_c(s)\), and hence
\(\eta_E(s)\simeq\eta_{\rm ch}\), when the TLS population is stationary on
the piston time scale.

The useful-work efficiency is obtained by replacing the total piston-energy
gain with the ergotropy gain,
\begin{equation}
\eta_{\mathcal W}(s)
=
\frac{\dot{\WP}(s)}
{\dot Q_h(s)}
=
\chi_E(s)\eta_E(s),
\qquad
\chi_E(s)
=
\frac{\dot{\WP}(s)}
{\dot E_P(s)}.
\label{eq:app_engine_ergotropy_efficiency}
\end{equation}
The factor \(\chi_E(s)\) accounts for the division of the piston-energy change
between ergotropy and passive energy and therefore depends on the evolving
piston state.

For the initially coherent piston considered in
Sec.~\ref{subsec:piston_amplification_ergotropy}, this state dependence can be
made explicit. Equations~\eqref{eq:AZD_ergotropy_growth_gain_form} and
\eqref{eq:piston_number_dynamics}, together with
\(E_P(s)=\hbar\nup n_P(s)\) and \(\Lambda(s)=-\Gamma(s)\), give
\begin{equation}
\chi_E^{\rm coh}(s)
=
\frac{
\Lambda(s)|\alpha(s)|^2
}{
D_P(s)+\Lambda(s)n_P(s)
}.
\label{eq:app_coherent_work_fraction}
\end{equation}
Here the numerator is the rate of coherent displacement-energy growth, which
equals the coherent-state ergotropy growth, whereas the denominator is the
total piston-energy growth and also contains the phase-insensitive excitation
generated by \(D_P(s)\). The coherent-state useful-work efficiency therefore
takes the form
\begin{equation}
\eta_{\mathcal W}^{\rm coh}(s)
=
\eta_E(s)
\frac{
\Lambda(s)|\alpha(s)|^2
}{
D_P(s)+\Lambda(s)n_P(s)
}.
\label{eq:app_coherent_ergotropy_efficiency}
\end{equation}
This expression retains the finite-time TLS-energy contribution through
\(\eta_E(s)\). When the TLS population is approximately stationary,
\(\eta_E(s)\simeq\eta_{\rm ch}\), and
Eq.~\eqref{eq:app_coherent_ergotropy_efficiency} reduces to
\begin{equation}
\eta_{\mathcal W}^{\rm coh}(s)
\simeq
\frac{\nup}{\omz}
\frac{
\Lambda(s)|\alpha(s)|^2
}{
D_P(s)+\Lambda(s)n_P(s)
}.
\label{eq:app_coherent_ergotropy_efficiency_stationary}
\end{equation}
Thus, in this regime, the coherent-state useful-work efficiency is the
channel-energy efficiency multiplied by the fraction of the piston-energy
growth stored as coherent ergotropy. Finite-time reservoir sampling can modify
this state-dependent fraction through the gain and excitation coefficients,
while the channel factor \(\nup/\omz\) remains fixed by the retained
transition energies. More general quantized-piston performance bounds that
explicitly account for piston entropy production and effective temperature
are given in Ref.~\cite{GelbwaserKurizkiPRE2014}.

For refrigerator operation, the retained cycle is reversed and
\(\Phi_c(s)<0\) whenever heat is extracted from the cold reservoir. From
Eqs.~\eqref{eq:refrigerator_cold_current} and
\eqref{eq:app_channel_energy_currents},
\[
\Jc(s)
=
\dot Q_c(s)
=
-\hbar\omm\,\Phi_c(s),
\qquad
-\dot E_P(s)
=
-\hbar\nup\,\Phi_c(s).
\]
The cold heat extracted and the piston energy consumed are therefore generated
by the same lower-sideband flux, giving
\begin{equation}
\varepsilon_{\rm ch}
=
\frac{\Jc(s)}
{-\dot E_P(s)}
=
\frac{\omm}{\nup}
=
\frac{\omm}{\omz-\omm}.
\label{eq:app_refrigerator_channel_cop}
\end{equation}
This frequency-ratio form is consistent with the autonomous
quantized-piston refrigerator framework of
Ref.~\cite{GelbwaserKurizkiPRE2014}. In contrast to the engine ratio in
Eq.~\eqref{eq:app_engine_operational_energy_efficiency},
Eq.~\eqref{eq:app_refrigerator_channel_cop} does not require
\(\dot p_e(s)\simeq0\), because its numerator and denominator are generated
by the same cold-sideband flux. The separate finite-resource condition for
positive cooling current is given by
Eq.~\eqref{eq:refrigerator_threshold_occupation}.

The piston resource can alternatively be characterized by the rate at which
its extractable work capacity is consumed. Whenever
\(\Jc(s)>0\) and \(\dot{\WP}(s)<0\), we define the ergotropy-based cooling
ratio
\begin{equation}
\varepsilon_{\mathcal W}(s)
=
\frac{\Jc(s)}
{-\dot{\WP}(s)}
=
\chi_R(s)\varepsilon_{\rm ch},
\qquad
\chi_R(s)
=
\frac{-\dot E_P(s)}
{-\dot{\WP}(s)}.
\label{eq:app_refrigerator_ergotropy_cop}
\end{equation}
The factor \(\chi_R(s)\) depends on the piston state through the energy
partition in Eq.~\eqref{eq:app_piston_energy_ergotropy_split}. Accordingly,
\(\varepsilon_{\mathcal W}(s)\) should be distinguished from the
channel-energy coefficient of performance
\(\varepsilon_{\rm ch}\). The more general thermodynamic COP bounds for a
quantized piston, including piston entropy production and effective
temperature, were derived in Ref.~\cite{GelbwaserKurizkiPRE2014}.

We finally relate the channel-energy ratios to the thermal operating
conditions. The exact KMS relation for the reservoir spectrum is given in
Eq.~\eqref{eq:app_detailed_balance}. In the Markovian limit,
\(\gamma_j(\omega)\rightarrow2\pi G_j(\omega)\), so the corresponding rate
ratio obeys detailed balance exactly. At finite elapsed time, however, each
rate samples a finite spectral interval through
Eq.~\eqref{eq:finite_time_rate_main}, and exact detailed balance for the rates
does not follow in general. When the thermal factor varies weakly over the
spectral range sampled by the finite-time kernel, the rates approximately
inherit the local thermal relation
\begin{equation}
\frac{\gamma_j(-\omega,s)}
{\gamma_j(\omega,s)}
\simeq
e^{-\beta_j\hbar\omega}.
\label{eq:app_finite_time_local_detailed_balance}
\end{equation}
An analogous finite-time detailed-balance regime underlies the anti-Zeno
enhancement of driven heat machines in
Ref.~\cite{MukherjeeCommunPhys2020}, where enhanced heat currents and power
can occur without a corresponding change in the channel-based efficiency or
coefficient of performance under the spectral conditions considered there.
Using Eq.~\eqref{eq:app_finite_time_local_detailed_balance}, the forward
engine-bias condition
\(
r_h^\uparrow r_c^\downarrow
\gtrsim
r_h^\downarrow r_c^\uparrow
\)
gives
\begin{equation}
\frac{\omm}{\omz}
\gtrsim
\frac{T_c}{T_h}.
\label{eq:app_engine_carnot_condition}
\end{equation}
Together with Eq.~\eqref{eq:app_engine_channel_efficiency}, this yields
\begin{equation}
\eta_{\rm ch}
\lesssim
1-\frac{T_c}{T_h}
\equiv
\eta_{\rm C}.
\label{eq:app_engine_channel_carnot_bound}
\end{equation}
For the reversed refrigerator bias,
\begin{equation}
\frac{\omm}{\omz}
\lesssim
\frac{T_c}{T_h},
\label{eq:app_refrigerator_carnot_condition}
\end{equation}
and Eq.~\eqref{eq:app_refrigerator_channel_cop} gives
\begin{equation}
\varepsilon_{\rm ch}
\lesssim
\frac{T_c}{T_h-T_c}
\equiv
\varepsilon_{\rm C}.
\label{eq:app_refrigerator_channel_carnot_bound}
\end{equation}
In the Markovian thermal limit these channel inequalities follow from exact
detailed balance; at finite time they remain accurate only over intervals in
which Eq.~\eqref{eq:app_finite_time_local_detailed_balance} is satisfied.
These are channel-energy bounds rather than universal bounds on
state-dependent ergotropy-based performance of a quantized piston
\cite{GelbwaserKurizkiPRE2014}.

Finite-time anti-Zeno sampling therefore modifies the retained transition
rates and event fluxes without changing the carrier and sideband energies.
The resulting channel-energy ratios remain fixed by the frequency geometry,
while useful-work performance additionally depends on the evolving piston
state through its ergotropy and passive-energy content.


\bibliography{references}

\end{document}